\documentclass{emulateapj}
\slugcomment{{\sc Accepted to ApJ} }%March 30 2013}

    \def\cm{\mathop{\rm cm}\nolimits}

  \def\etal{{\it et al }}
\def\eg{{\it e.g.\ }}   \def\ie{{\it i.e.}}
   
   \def\p3m{P${}^3$M}
\def\ap3m{AP${}^3$M} \def\-{{\em{---}}} \def\msun{{M_\odot}}

\newcommand{\be}{\begin{equation}}
\newcommand{\ee}{\end{equation}}
\def\rs{{$r_{\rm s}$}}
\def\tff{$\tau_{\rm ff}$ }

\def\hp{{H$^{+}$}}
\def\hm{{H$^{-}$}} 
\def\hep{{He$^{+}$}}
\def\cp{{C$^+$}} 
\def\cm{{C$^{-}$}}  
\def\op{{O$^+$}} 
\def\om{{O$^{-}$}} 
\def\htwo{{H$_{2}$}} 
\def\htwop{{H$_{2}^{+}$}} 
\def\hthreep{{H$_{3}^{+}$}}
\def\ct{{C$_2$}} 
\def\ot{{O$_2$}} 
\def\otp{{O$_2^+$}} 
\def\oh{{OH}} 
\def\ohp{{OH$^+$}} 
\def\co{{CO}} 
\def\cop{{CO$^+$}}
\def\ch{{CH}} 
\def\chp{{CH$^+$}}
\def\cht{{CH${_2}$}} 
\def\chtp{{CH${_2^+}$}} 
\def\chthp{{CH$_3^+$}} 
\def\hcop{{HCO$^+$}} 
\def\hocp{{HOC$^+$}}
\def\hto{{H$_2$O}} 
\def\htop{{H$_2$O$^+$}} 
\def\hthop{{H$_3$O$^+$}}
\def\elec{{e$^{-}$}}

\usepackage{epstopdf} \usepackage{subfigure} \usepackage{rotating} \usepackage{amsmath}

\begin{document}

\title{Thermal and Chemical Evolution of Collapsing Filaments}
\author{{William J. Gray\altaffilmark{1} \&  Evan Scannapieco}\altaffilmark{2}}
\altaffiltext{1}{Lawrence Livermore National Laboratory, P.O. Box 808, L-038, Livermore, CA, 94550.}
\altaffiltext{2} {School of Earth and Space Exploration, Arizona State University, P.O. Box 871404, Tempe, AZ, 85287-1494.}

\begin{abstract}

Intergalactic filaments form the foundation of the cosmic web that connect galaxies together, and provide an important reservoir of gas for galaxy growth and accretion. Here we present very high resolution two-dimensional simulations of the thermal and chemical evolution of such filaments, making use of a 32 species chemistry network that tracks the evolution of key molecules formed from hydrogen, oxygen, and carbon.  We study the evolution of filaments over a wide range of parameters including the initial density, initial temperature, strength of the dissociating UV background, and metallicity. In low-redshift,  $Z \approx 0.1 Z_\odot $ filaments,  the evolution is determined completely by the initial cooling time.  If this is sufficiently short,  the center of the filament always collapses to form dense, cold core containing a substantial fraction of molecules.  In high-redshift,   $Z=10^{-3} Z_\odot$  filaments, the collapse proceeds much more slowly.  This is  due mostly to the lower initial temperatures, which leads to a much more modest increase in density before the atomic cooling limit is reached, making subsequent molecular cooling much less efficient.

Finally, we study how the gravitational potential from a nearby dwarf galaxy affects the collapse of the filament and compare this to NGC 5253, a nearby starbusting dwarf galaxy thought to be fueled by the accretion of filament gas. In contrast to our fiducial case, a substantial density peak forms at the center of the potential. This peak evolves faster than the rest of the filament due to the increased rate at which chemical species form and cooling occur. We find that we achieve similar accretion rates as NGC 5253 but our two-dimensional simulations do not recover the formation of the giant molecular clouds that are seen in radio observations. 

\end{abstract}

\section{Introduction}

In the modern theory of cosmic evolution, structures formed from the bottom up through continuous accretion and mergers (\eg White \& Rees 1978; White \& Frenk 1991; Kauffmann \etal 1993; Cole \etal 2000; Bower \etal 2006). Originating as small perturbations in the initial density field, seed overdensities were slowly transformed by non-linear growth into the complex distribution of galaxies observed in large scale surveys (\eg Doroshkevich \etal 2004;  Ratcliffe \etal 1996;  Pandey \& Bharadwaj 2006) and seen in $N$-body simulations (\eg Bond, Kofman, \& Pogosyan 1996). Between these growing structures is a complex network of filaments, which connects them together and plays a key factor in their evolution. 

This cosmic web provides a reservoir of gas by which galaxies grow and accrete (White \& Rees 1978), which may explain the nearly constant star-formation experienced by the Milky Way for example (Binney, Dehnen \& Bertelli 2000). The state of the gas flowing along a filament is predominately determined by the mass of the dark matter halo (\eg Birnboim \& Dekel 2003; Dekel \& Birnboim 2006; Kere$\breve{\rm s}$ \etal 2005). Traditionally, two modes of accretion are discussed. ``Hot mode'' accretion occurs when the inflowing gas is shock heated as it is compressed by the hot gas of the halo and typically occurs in galaxies with masses above 10$^{12}\msun$ (Thoul \& Weinberg 1995; Furlanetto \& Loeb 2004; Birnboim \& Dekel 2003; Dekel \& Birnboim 2006). Below this threshold, the gas is not shock heated and accretes directly onto the galaxy. At high redshifts, the cold mode dominates as most galaxies are not massive enough to shock heat the infalling gas. Conversely, the hot mode dominates at low redshift ($z<$ 2) where most massive galaxies form (Kacprzak \etal 2011; Stewart \etal 2011).

In addition, filamentary accretion may help explain the formation of the observed cold neutral hydrogen clouds that surround the Milky Way. These high-velocity clouds (HVCs) could form in the Milky Way halo from a variety of mechanisms: cooling instabilities in the hot circumgalactic medium (Mo \& Miralda-Ecud$\acute{\rm e}$ 1996; Maller \& Bullock 2004; Joung \etal 2012),  production from a galactic fountain where hot gas is driven from the disk, cools and condenses, and rains back onto the disk (Bregman 1980; Melioli \etal 2008, 2009; Marinacii \etal 2011), or the interaction between hot halo gas and an infalling satellite (\eg Putman \etal 2003). On the other hand, Kere$\breve{\rm s}$ \etal (2009) suggest that through turbulent fragmentation, the filament gas itself is able to create these cold clouds. 

Observing filaments has long been a challenging task. Direct detection attempts initially concentrated on X-ray surveys, motivated by the fact that most of the filamentary gas is expected to be very hot (\eg Pierre, Bryan, \& Gastaud 2000). Only recently have such studies been successful (Werner \etal 2008; Walker \etal 2012; Williams \etal 2010). Weak lensing, on the other hand, have has some success in studying filaments of galaxy cluster pairs (Gray \etal 2002; Dietrich \etal 2005). Still others have derived filament properties using a variety of statistical or topological models of large scale surveys. These procedures have estimated the width of filaments to be 1-2 Mpc (Ratcliffe \etal 1996; Doroshkevich \etal 2004) and tens of Mpc in length (Bharadwaj, Bhavsar \& Sheth 2004). 

With these limited direct constraints, $N$-body simulations have played an instrumental role in our understanding of the growth of the cosmic web.  Dolag \etal (2006) studied such a simulation and found that many of the filaments have density profiles that scale as $r^{-2}$-$r^{-3}$ with temperatures between $\sim$10$^{5}$-10$^{7}$ K. Similar results are found by Colberg \etal (2005). Harford \etal (2011) studied a similar filament at $z\sim 5$ and found slightly lower temperature $1-2\times10^{4}$ K, but with similar central densities of a few 10-1000 times the average background density.

 While star formation in a Milky Way size galaxy is relatively constant  it is a different matter for dwarf galaxies. The star formation per unit mass in these galaxies is just as intense as that seen in larger spiral galaxies without the expected dynamical causes, such as spiral arms and bars (\eg Vanzi \etal 2009). The star formation is episodic and driven by mechanisms that are poorly understood, although mergers, tidal interactions, and accretion from filaments are possible causes (\eg Cumming \etal 2008; James \etal 2010). In addition, the metallicity of these galaxies is lower relative to larger star forming galaxies and they may provide insights to star formation during earlier formation epochs (Meier \etal 2002). 

NGC 5253 represents a prime example of such a galaxy. It is a companion galaxy to M83 and is nearby at a distance of 3.8 Mpc (Gibson \etal 2000). Recent observations have uncovered multiple ``super-star clusters'', which are very bright with $L_{IR}\sim 10^9 L_{\odot}$ (\eg Gorjian, Turner, \& Beck 2001) and $M_V \sim$ -10 to -11 (Meurer \etal 1995; Gorjian 1996). In addition, radio observations of CO(2-1) emission coincident with a dust lane and with properties consistent with inflow (Meier \etal 2002). However, the nature of the starburst and how large a part the infall plays remains an open question. 

It is with these observations in mind that we present the results from a series of 2D numerical simulations that explore the evolution of an idealized intergalactic filament. We concentrate on the chemical and molecular makeup of the filament as it evolves from an initially atomic ionized state. At high temperatures, the cooling is dominated by atomic species. At lower temperatures, on the other hand, cooling from molecules and metals is more important. Here we track the formation and evolution of the key moleuclar species H$_2$, CO, OH, and H$_2$O, that contribute most to the cooling at low temperatures. These results then allow us to asses the ability for filaments to cool to low temperatures and fragment into dense clouds.  Finally, we extend our simulations to model a filament near a dwarf galaxy, comparable to NGC 5253.

The structure of this paper is as follows. In \S2 we introduce our chemistry and cooling routines and discuss the implementation and associated tests. In \S3 we outline our filament model 
and in \S4  we present our results for isolated filaments.   In \S5 we present results for filaments forming near a dwarf galaxy, and a discussion and conclusions are given in \S6. 

\section{Numerical Method}

All simulations were performed with FLASH version 3.3, a publicly-available multidimensional adaptive mesh refinement hydrodynamic code (Fryxell \etal 2000) that solves the Riemann problem on a Cartesian grid using a directionally-split piecewise parabolic method (PPM) (Colella \& Woodward 1984; Colella \& Glaz 1985; Fryxell 1990).  Two new capabilities needed to be added to the public version of FLASH to properly simulate our filaments: a non-equilibrium chemistry package that tracks the formation and evolution of important molecular coolants, and a cooling routine that takes into account the cooling from each of these molecules. In this section we describe the numerical implementation and tests for each of these new capabilities. 

\subsection{Chemistry}

	Even when only a few elements are considered,  the gas phase chemistry of enriched gas can be quite complex, with hundreds of species and thousands of individual reactions  (\eg Le Teuff \etal 2010; Semenov \etal 2010). While such large networks are important for many studies of the interstellar medium (ISM), they are impractical when coupled with large hydrodynamical simulations. We therefore used a substantially smaller network that tracks and evolves the atomic and molecular species made up of the elements that are most important in the thermodynamics of our system: hydrogen, helium, carbon, and oxygen.
	
\subsubsection{Implementation}

	In particular, we have implemented the chemical network presented in Glover \etal (2010), which tracks the impact of 218 separate chemical reactions on 32 species: atomic hydrogen (H, \hp, \hm), atomic helium (He, \hep), atomic carbon (C, \cp, \cm), atomic oxygen (O, \op, \om), molecular hydrogen (\htwo, \htwop, \hthreep), molecular carbon (\ct), molecular oxygen (\ot, \otp), key molecules containing combinations of these elements (\oh, \ohp, \co, \cop, \ch, \chp, \cht, \chtp,\chthp, \hcop, \hocp, \hto, \htop, \hthop), and free electrons (\elec).  As the binding and ionization energies of each species are important for the overall energy budget of the gas, we assigned each species with a energy, $E_v \equiv B.E. - I.E,$ which is simply the ionization potential subtracted from the binding energy, as summarized in Table~\ref{senergy}.  	
	
	In addition to gas phase collision reactions, our network included the effect of an ambient ultraviolet (UV) radiation field. The photochemical rates assume a standard interstellar radiation field from Draine (1978) which has a field strength of $G_0$ = 1.7 in Habing (1968) units.  This corresponds to 6.5$\times10^{-4}$ FUV photons per cm$^{-3}$. We introduced a coefficient, J$_{21} \equiv G/G_0$ that allowed us to alter this background field, assuming that the field scales linearly with $G_0$. For example, a J$_{21} = 0$ removes the background field while a value of J$_{21} = 1.0$ gives the standard Habing field.  
%Similarly, we used the parameter $\xi_H$ to alter the background cosmic ray ionization rate. Finally, many of these rates are dependent  on the visual extinction between the ionizing source and position of the % gas, and for simplicity we treated this extinction as another free variable throughout the simulation volume.  
	
\begin{table}
\caption{Summary of species data. The first column gives the species name, the second column gives the ionization potential, third gives the dissociation energy, fourth gives E$_v$, and the last column gives the ratio of specific heats. All energies are given in units of eV. Those species denoted with a single asterisk do not have published dissociation energies. Finally, \hocp,  which we denote with the double asterisk, does not have any published ionization or dissociation energies, and we assume they are the same as for \hcop.}
\begin{centering}
\begin{tabular}{llllcl}
\hline
     Species & I.P (eV) & D.E (eV) &   E$_v$ (eV) &  $\gamma$ &            \\
  \hline
      H 	&	0.00 		&       0.00 &       0.00 		&       1.66 &            \\
      \hp 	&       13.60 	&       0.00 &     -13.60 	&       1.66 &            \\
      \hm 	&       0.00 		&       0.77 &       0.77 		&       1.66 &            \\
      He 	&       0.00 		&       0.00 &       0.00 		&       1.66 &            \\
      \hep 	&       24.60 	&       0.00 &     -24.60 	&       1.66 &            \\
     C 		&       0.00 		&       0.00 &       0.00 		&       1.66 &            \\
     \cp 	&       11.27 	&       0.00 &     -11.27 	&       1.66 &            \\
     \cm 	&       0.00 		&       1.26 &       1.26 		&       1.66 &            \\
     O 		&       0.00 		&       0.00 &       0.00 		&       1.66 &            \\
     \op 	&       13.60 	&       0.00 &     -13.60	&       1.66 &            \\
     \om 	&       0.00 		&       1.46 &       1.46 		&       1.66 &            \\
     \htwo 	&       0.00 		&       4.48 &       4.48 		&       1.40 &            \\
     \htwop &       15.43 	&       4.48 &     -10.95 	&       1.40 &            \\
     \hthreep &   	16.30 	&       6.30 &     -10.00 	&       1.31 &            \\
     \ct 	&       0.00 		&       6.21 &       6.21 		&       1.40 &            \\
     \ot 	&       0.00 		&       5.12 &       5.12 		&       1.40 &            \\
     \otp 	&       12.06	&       6.66 &      -5.40 	&       1.40 &            \\
     \oh 	&       0.00 		&       4.39 &       4.39 		&       1.40 &            \\
     \ohp 	&       13.00 	&       5.10 &      -7.90 	&       1.40 &            \\
     \co 	&       0.00 		&      11.90 &      11.90 	&       1.40 &            \\
     \cop 	&       14.01 	&       8.34 &      -5.67 	&       1.40 &            \\
     \ch 	&       0.00 		&       3.47 &       3.47 		&       1.40 &            \\
     \chp 	&       10.64 	&       0.00 &     -10.64 	&       1.40 &          * \\
     \cht 	&       0.00 		&       4.00 &       4.00 		&       1.31 &            \\
     \chtp 	&       10.40 	&       4.00 &      -6.40 	&       1.31 &            \\
     \chthp 	&       9.83 		&       0.00 &      -9.83 	&       1.31 &          * \\
     \hcop 	&       9.88 		&       0.00 &      -9.88 	&       1.31 &          * \\
     \hocp 	&       9.88 		&       0.00 &      -9.88 	&       1.31 &         ** \\
     \hto 	&       0.00 		&       0.00 &       0.00 		&       1.31 &            \\
     \htop 	&       12.61 	&       5.11 &      -7.50 	&       1.31 &            \\
     \hthop 	&       0.00 		&       0.00 &       0.00 		&       1.31 &          * \\
     \elec 	&       0.00 		&       0.00 &       0.00 		&       1.66 &            \\
\label{senergy}
\end{tabular}  
\end{centering}
\end{table}

Here we briefly describe our implementation of this network.  First, we labele each species by an index $i,$  such that species $i$ had $Z_i$ protons, $A_i$ nucleons, and a mass density of $\rho_i.$ From this we define a mass fraction as $X_i \equiv \rho_i / \rho,$  where $\rho = \sum_i \rho_i$ is the overall gas density, and define the molar mass fraction as $Y_i \equiv X_i/A_i.$  A continuity equation for each species is then given as $\dot{Y}_i =  \dot{R}_i,$ where $\dot{R}_i$ is the total reaction rate due to all reactions. We then follow the evolution of every species in our simulation except for free electrons which we directly calculate using conservation of the total electron density.

Due to the complex ways that chemical reaction rates depend on temperature and the wide range of abundances possible, the resulting rate equations are `stiff'. This requires implicit or semi-implicit methods to properly solve these equations.  To that end,  we have implemented a fourth-order accurate Kaps-Rentrop, or Rosenbrock method (Kaps \& Rentrop 1979), as described in Gray \& Scannapieco (2010; hereafter GS10). As the species evolve, the temperature will change as the internal energy is altered from recombinations, ionizations, and dissociations. Since the reaction rates are strong functions of temperature, the network can become unstable if too large of a time step is used. To ensure the stability of the network while running the simulation at the hydrodynamic time step, we use the same subcycling method from GS10. A chemical time step was calculated for each species as 
\be
\tau_{{\rm chem},i} = \alpha_{\rm chem} \frac{Y_i + 0.1Y_{H^+}}{\dot{Y_i}},
\ee
where $\alpha_{\rm chem}$ is a constant determined at runtime and controls the maximum abundance change allowed, which we default to 0.1.  We used the current value for the species abundances and the change in abundances ($\dot{Y_i}$) were calculated from the ordinary differential equations at the current temperature.  Finally, we included a small fraction of $Y_{H^+}$  as a buffer to prevent species with very small abundances, but changing  rapidly, from causing prohibitively small time scales and therefore excessive subcycling. 

At the beginning of the chemistry cycle, the smallest species time scale was compared to the hydrodyanmic time step. If the hydro time is smaller than the chemistry time step, no additional subcycling is done and the network was advanced as normal. If the chemistry time step is smaller than the hydro time step, then the network is advanced using the chemistry time step. The newly updated species and temperature are used to recalculate the chemistry time step, which was then compared to the remaining hydro time. This continues until the species have advanced a full hydro time step.

Finally, there are two special cases where the chemistry can be simplified. At temperatures greater than 1.7$\times$10$^{7}$ K, we assumed that all species are atomic and ionized and no additional chemistry is done. At temperatures greater than 1.0$\times$10$^{4}$ K but less than 1.7$\times$10$^{7}$ K, we assume that all molecules are dissociated and added to the atomic neutral species (\eg X$_{\rm H}$ = X$_{\rm H}$ + X$_{\rm H_2}$ + ...). A chemistry step is then performed.

\subsubsection{Chemistry Test}

As a test of our implementation of  these coupled equations, we used a procedure similar to that presented in GS10 and Glover \etal (2010).  In order to have a commensurate dataset to compare against, we use reaction rates given in the UMIST RATE06 database (Woodall \etal 2007), and restricted our attention to two-body interactions in the absence of a background UV dissociating field. Out of the 4604 reactions in RATE06, 134 were used that matched the reactions given in Glover \etal (2010). 

We completed this complementary network by adding rates that appear in Glover \etal (2010) and not in RATE06. In total, this gave a total of 144 reactions. In particular we added reactions for the positive-negative ion recombination of \om\ and \hp\ (Rxn 108 in Glover et al 2010), the formation of \hthop\ from \htwo\ and \hp\ (Rxn 141), the formation of \htwop\ from \hm\ and \hp\ (Rxn 15), the formation of \htwop\ from H and \hp\ (Rxn 4), the charge exchange reaction between \htwo\ and \hp\ (Rxn 7), ion molecule reactions between \hthop\ and H (Rxn 55), between \op\ and \htwo\ (Rxn 69), and between \co\ and \hep\ (Rxn 105), and the electron attachment of \hcop\ (Rxn 130). 

Next, we compare the reaction rates presented in Glover \etal (2010) to those in RATE06. When these rates are appreciably different, we alter those rates to match those found in RATE06. Specifically, we changed the dissociation of \htwo\ by H (Rxn 9) and of \htwo\ by \htwo\ (Rxn 10), the radiative recombination of \hp\ (Rxn 12) and \hep\ (Rxn 17), the formation of \co\ by \ch\ and O (Rnx 38), the formation of \ot\ by \oh\ and O (Rxn 47), the formation of \hthreep\ by \htwop\ and \htwo\ (Rxn 54), the dissociation of \co\ by \hep\ (Rxn 104), the dissociative recombination of \hthreep\ (Rxn 110-112), the dissociative recombination of \htwop\ (Rxn 120-122), and the dissociative recombination of \hthop\ (Rxn 123-126). 

Having matched our FLASH network with a restricted set of RATE06 reactions, we then ran a series of simple tests. We use a series of single zone models to compare the chemical evolution over a wide range of intital temperatures and densities. In each model, the thermodynamic variables are kept constant. To advance the RATE06 reaction network, we use the open source chemistry code {\it Astrochem}\footnote{http://smaret.github.com/astrochem/}. Models were run for a total of 10$^8$ years with hydrogen number densities ranging between 10$^{-2}$ to 10$^{6}$ cm$^{-3}$ and temperatures ranging between 10$^2$ to 10$^4$ K. In all cases the initial abundances of each species relative to the total hydrogen density were: $n_{\rm He}$ = 0.08, $n_{\rm C^+}$= 8.2$\times$ 10$^{-5}$, $n_{\rm O^+}$ = 1.5$\times$ 10$^{-5}$, $n_{\rm H^+}$ = 0.99, and $n_{\rm H}$ = 0.01.  We then ran similar models with our modified FLASH code, starting each model with a small initial time step ($\sim$ 10$^5$ s) which was allowed  to increase to the hydrodynamic time step. 

\begin{figure}
\centering
\includegraphics[trim=0.0mm 0.0mm 0.0mm 0.0mm,clip,scale=0.45]{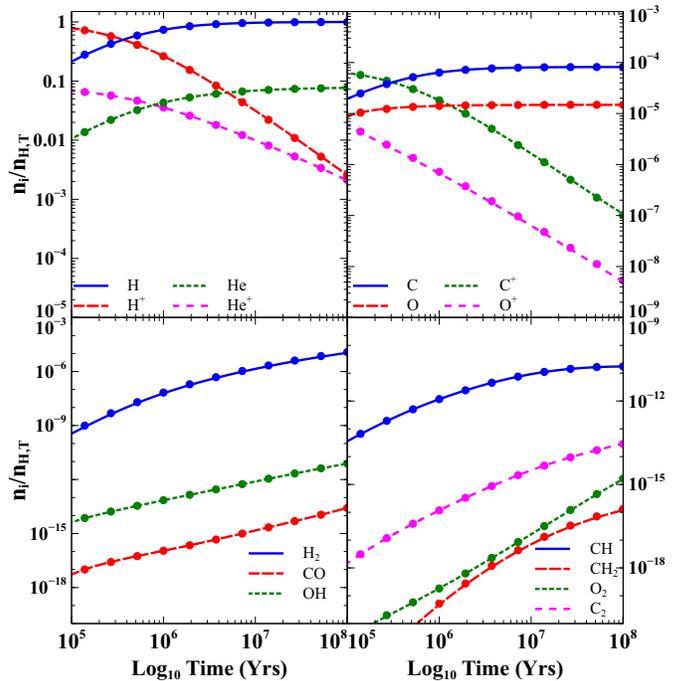}
\caption{Comparison of chemical abundances between {\it Astrochem} and FLASH. In each panel the lines show the abundances from FLASH while the circles are from {\it Astrochem}.  The $y$-axis of each panel is the logarithm of time in units of years and the number density of the given species divided by the total number density of hydrogen. {\it Top Right:} The dashed (red) lines show \hp, the solid (blue) lines show H, the fine-dashed (magenta) lines show \hep , and the dotted (green) lines show He. {\it Top Left:} The dashed (red) lines show O, the solid (blue) lines show C, the fine-dashed (magenta) lines show \op, and the dotted (green) lines show \cp. {\it Bottom Right:} The dashed (red) lines show \htwo, the solid (blue) lines show \co, and the dotted (green) lines show \oh. {\it Bottom Left:} The dashed (red) lines show \ch, the solid (blue) lines show \cht, the dotted (green) lines show \ot, and the fine-dashed (magenta) lines show \ct. In general, the solid lines and circles always overlap each other and it is very hard to differentiate between them.  }
\label{ChemTest}
\end{figure}

Figure \ref{ChemTest} shows a comparison between the derived chemistry network evolved with {\it Astrochem} and the FLASH network where the temperature is constant at 100 K and a hydrogen number density of $n_{\rm H}$ = 10$^{-2}$ cm$^{-3}$. While the abundances change over many orders of magnitude through the evolution time, our FLASH always closely match the {\it Astrochem} network. Similar results are found for the range of conditions considered. 
% At the temperature considered here, the ionized species very quickly begin to recombine into their neutral state. However, even during this quick period a substantial amount of molecular species are created. At both early and late times both the FLASH and {\it Astrochem} agree very well and are almost indistinguishable.

Another simple test was to ensure that the chemical abundances scaled appropriately with density. The timescale for formation of a given chemical species can be estimated as $\tau_{\rm chem}\sim n/\dot{n}$, where $n$ and $\dot{n}$ are the instantaneous species number density and its rate of change, respectively. At low densities where three body reactions are unimportant, $\dot{n}$ is given by $k n^2$, where $k$ is the formation or destruction rate, a function of only temperature. This gives a chemistry timescale that is inversely proportional to the overall density. We ran several tests in which where we compared the time evolution of our chemical species at a constant temperature and changed the initial mass density. Although not shown here, the species always evolved as expected. 

\subsection{Cooling} 

As the species evolve, the temperature of the gas changes due to recombinations, ionizations, and dissociations, as described above. In addition, at low temperatures (>10$^4$ K), the cooling from molecular species becomes important. Since both the reaction and cooling rates are temperature dependent, it is important for both of these processes to remain coupled. Therefore, once a chemical subcycle was performed, the cooling routines were immediately called. This ensured the stability of both the chemistry and cooling routines.

At temperatures $\ge$10$^{4}$ K, the majority of radiative cooling comes from atomic species, with bremmstrahlung radiation becoming important at temperatures $\ge$10$^{7}$K. Below 10$^{4}$K, many cooling channels become important, primarily from atomic metal lines from carbon and oxygen, as well as molecular line cooling from \htwo, \co, \hto, and \oh. In particular, carbon monoxide (CO) is extremely important at temperatures below 200 K where molecular hydrogen (\htwo) becomes inefficient (Smith \etal 2008).   Finally, as in Glover \etal (2010), we included heating from the photoelectric effect, H$_2$ UV pumping.%, and cosmic ray ionization.

In this case, the total cooling function becomes,
\be
\Lambda_{\rm Total} = \Lambda_{\rm H,He} + \Lambda_{\rm Molecules} + \Lambda_{\rm Metals} - \Lambda_{\rm Heating},
\label{CF}
\ee
such that the total cooling function in units of ergs s$^{-1}$ cm$^{-3},$ is simply the summation of the cooling rate from each cooling channel minus sources of gas heating. The introduction of carbon and oxygen into the chemistry network allows for the creation of many molecules that are  important for the thermal evolution of the gas. As we have expanded the chemistry network, we also expanded the cooling routine presented in GS10 to include these important coolants. Much of this cooling follows from Glover \etal (2010), which we summarize below. 

\subsubsection{Atomic Cooling}
The treatment of the high temperature hydrogen and helium cooling remained unchanged from GS10. The cooling rate is calculated using CLOUDY (Ferland \etal 1998) assuming only collisional ionization in the Case B limit. 

To account for metal-line cooling, we used the tabulated results from Weirsma \etal (2008), assuming standard solar rations scaled by the local metallicity.  These rates assume local thermodynamic equilibrium and are defined from 10$^2$ to 10$^9$ K.   Note that while cooling by atomic hydrogen and helium are important above 10$^4$K, below this temperature atomic cooling comes primarily from neutral oxygen and  neutral and singly ionized carbon.

\subsubsection{Molecular Cooling} 

%\subsubsubsection{\htwo\ cooling}
\paragraph{\htwo\ cooling}

Much of the \htwo\ cooling rates are taken from Glover \etal (2008) and again remain unchanged from GS10. Each cooling rate has the form:
\be 
\Lambda_{i,j} = n_i n_j \lambda_{i,j},
\label{mcf}
\ee
where $\Lambda_{i,j}$ is the energy loss per volume due to species $i$ and $j$, $n_i$ and $n_j$ are the number densities of each species, and $\lambda_{i,j}$ is the cooling rate in ergs cm$^{-3}$ s$^{-1}$. Cooling rates for the collisional excitation between \htwo\ and H, \htwo , \hp , and \elec\ and between \htwop\ and H or \elec, are all included. 

%\subsubsubsection{\co\ and \hto\ cooling}
\paragraph{\co\ and \hto\ cooling}

To implement \co\ and \hto\ cooling, we followed Glover \etal (2010).  Here the authors used the tabulated rotational cooling functions from Neufeld \& Kaufman (1993) and Neufeld, Lepp, \& Melnick (1995). As above, the cooling rate is defined as a rate coefficient $L$, with units of erg cm$^{3}$ s$^{-1}$ such that the cooling rate per unit volume per unit time is  $\Lambda$=$Ln(H_2)n(M)$, where $n(H_2)$ is the number density of \htwo\ and $n(M)$ is the number density of the other molecular species.

The cooling rate coefficient is given as a function of the \htwo\ number density, the kinetic temperature, and a parameter that relates the species number density to the local velocity field $\widetilde{N}(m)$, defined as:
\be
\widetilde{N}(m) = \frac{n(m)}{| \nabla \cdot {\rm v}|}.
\label{nm}
\ee
The rate coefficients for CO and ${\rm H_2O}$ cooling are computed as a four-parameter analytic fit of the form:
\be
\frac{1}{L} = \frac{1}{L_0} + \frac{n( {\rm H_2} )}{L_{\rm LTE}} + \frac{1}{L_0} \left( \frac{n(\rm H_2)}{n_{1/2}} \right)^{\alpha} \left( 1 - \frac{ n_{1/2} L_0}{L_{\rm LTE} } \right).
\label{lfun}
\ee
Here $L_0$ represents the cooling in the low density limit, $L_{\rm LTE}$ is the cooling rate when the rotational level populations are in local thermodynamic equilibrium, and $n_{1/2}$ is the \htwo\ number density when $L$ = 0.5$L_0$. $L_0$ is a function of temperature only while the other three parameters ($L_{\rm LTE}$, $n_{1/2}$, and $\alpha$) are functions of temperature and the effective column density per unit velocity as defined in eq.\ (\ref{nm}).

Rotational cooling for \hto\ the cooling rate coefficient was tabulated between 10 K $< {T} <$ 4000 K and for column densities between 10 $< \log_{10}(\widetilde{N}({\rm H_2O})) <$ 19, where $\widetilde{N}({\rm H_2O})$ in units of cm$^{-2}$ per km s$^{-1}$. We assumed a constant ortho-para ratio of 3:1 as Neufeld \etal (1995) who define cooling parameters for both states of \hto . Similarly, the rotation cooling of \co\ is defined for temperatures between 10 K $< T <$ 2000 K and for column densities between 14.5 $< \log_{10}(\widetilde{N}({\rm CO})) <$ 19. 

We also introduced an artificial temperature floor at 10 K and turn off any radiative cooling at these temperatures. To properly model gas to temperatures below this value, one would be required to not only extend the results from Neufeld \etal (1993; 1995) but also include other physical processes, such as the CO freeze-out (\eg Lee \etal 2004) or, if included in the chemical network, the decrease in dust temperature as the extinction increases (\eg Goldsmith 2001).   

For gas with temperatures above the tabulated values, we simply adopted cooling rates that correspond to the highest tabulated temperature. If the column density was within the tabulated range, then we linearly interpolated along column density to determine our cooling rate. Although at such high temperatures, it is expected that these coolants are quickly dissociated.  

A similar process was used when exceeding the tabulated column density. If a zone was denser than tabulated, we used the highest values given for the cooling rate and, if possible, interpolated along temperature. If a zone was more rarified than tabulated, then  we used the lowest tabulated values. At very high densities, this procedure may overestimate the expected cooling somewhat. Although, in our simulations the abundance of these coolants was negligible at such low densities, and they did not contribute much to the overall cooling. At lower densities however, the cooling rates naturally approach the optically thin limit and it is unlikely to introduce much error in our simulations.  

Finally, in their calculation of the cooling functions, Neufeld \etal (1993; 1995) assumed that only collisions between the target coolant and \htwo\ are important. However, as pointed out by Glover \etal (2010), if the gas is not completely molecular then the collisions between atomic hydrogen and electrons also contribute. To account for this, we replaced $n_{\rm H_2}$ in eq.~\ref{lfun} and in the total cooling rate, with an effective number density, $n_{\rm eff}$. For \co\ rotational cooling the effective number density is

\begin{eqnarray}
\small
n_{\rm eff,CO,rot} &=& n_{\rm H_2} + \sqrt{2}\left( \frac{\sigma_{\rm H}}{\sigma_{\rm H_2}}\right) n_{\rm H} \nonumber \\
&+& \left( \frac{1.3\times10^{-8} {\ \rm cm^3\ s^{-1}}} {\sigma_{\rm H_2} v_e} \right) n_{\rm e},
\end{eqnarray}
where $n_{\rm H_2}$, $n_{\rm H}$, and $n_{\rm e}$ are the number densities of \htwo , hydrogen, and electrons respectively, $\sigma_{\rm H}$ = 2.3$\times$ 10$^{-15}$ cm$^2$, $\sigma_{\rm H_2}$ = 3.3$\times$10$^{-16}$(T/1000K)$^{-1/4}$ cm$^2$, and $v_e$=1.03$\times$10$^4$ $\sqrt{T({\rm K})}$ cm s$^{-1}$. For \hto\ rotational cooling the effective number density is
\be
n_{\rm eff,H_2O,rot}=n_{\rm H_2} + 10 n_H + \left( \frac{k_e}{k_{\rm H_2}} \right) n_e,
\ee
where $k_e$ = dex[-8.020 + 15.740/$T^{1/6}$ - 47.137/$T^{1/3}$ + 76.648/$T^{1/2}$ - 60.191/$T^{2/3}$] and $k_{\rm H_2}$ = 7.4$\times$10$^{-12}\ T^{1/2}$ cm$^3$ s$^{-1}$. These equations are taken from Meijerink \& Spaans (2005) while the formula for $k_e$ is taken from Faure, Gorfinkiel \& Tennyson (2004). 
 
Vibrational cooling from \co\ and \hto\ , which is important at high temperatures and densities, is also presented in Neufeld \etal (1993). These authors provide a simpler two parameter fit of the form
\be
\frac{1}{L} = \frac{1}{L_0} + \frac{n_{\rm H_2}}{L_{LTE}}.
\ee
Analytical functions are provided for $L_0$ for \co\ and \hto\ as well as tabulated values for $L_{LTE}$. Vibrational cooling rates for both coolants are tabulated for temperatures 100 K $< T <$ 4,000 K and for column densities of 13 $< \log_{10} (\widetilde{N}) <$ 20. As above, $n_{\rm H_2}$ is replaced with updated values from Meijerink \& Spaans (2005). For \co\ vibrational cooling, the effective number density is
\be
n_{\rm eff,CO,vib} = n_{\rm H_2} + 50 n_{\rm H} + \left( \frac{L_{\rm CO,e}}{ L_{\rm CO,0}} \right) n_{\rm e},
\ee
where $L_{\rm CO,e} = 1.03 \times 10^{-10} \left( \frac{T}{300} \right)^{0.938}{\rm exp} \left( \frac{-3080}{T} \right)$, and $L_{\rm CO,0} = 1.14 \times 10^{-14}{\rm exp} \left( \frac{-68.0}{T^{1/3}} \right) {\rm exp} \left( \frac{-3080}{T} \right)$.
Similarly for \hto\ vibrational cooling, the effective number density is

\be
n_{\rm eff,H_2O,vib} = n_{\rm H_2} + 10 n_{\rm H} + \left( \frac{L_{\rm H_2O,e}}{L_{\rm H_2O,0}} \right) n_{\rm e},
\ee
where $L_{\rm H_2O,e} = 2.6 \times 10^{-6} T^{-1/2} {\rm exp} \left( \frac{-2325}{T} \right)$,
and $L_{\rm H_2O,0} = 0.64 \times 10^{-14} {\rm exp} \left( \frac{-47.5}{T^{1/3}} \right) {\rm exp} \left( \frac{-2325}{T} \right)$.
For temperatures and effective column densities outside of the tabulated range, we calculated cooling rates as we did for the rotational case. 

%\subsubsubsection{\oh\ cooling}
\paragraph{\oh\ cooling}

To model the cooling from \oh\ we used the tabulated results from Omukai \etal (2010), who computed the \oh\ cooling rate in the same manner as Neufeld \etal (1993; 1995) and presents values to fit eq.\ (\ref{lfun}). The rate is valid in a temperature range between 30 K $< T <$ 600 K and for effective column densities of 10 $< log_{10} ( \widetilde{N}({\rm OH}) ) < $ 18. For temperatures and densities outside this range, we calculated the appropriate cooling as above.  Here we assume that only direct collisions between OH and \htwo\ determine the cooling rate.  

%\subsubsubsection{Heating}
\paragraph{Heating}

To complete our treatment of the thermal evolution of the gas, we included heating by the photoelectric effect and \htwo\ UV pumping as modeled by Glover \etal (2010). We also included a rate that takes into account the photoionization of hydrogen. When hydrogen is collisionally ionized, energy is taken from the gas to overcome the binding energy of the atom and lowering the internal energy of the gas. However, if the atom is photoionized, it does not alter the internal energy. We model this as a heating term of the form:
\be
\Gamma_{\rm H^+} = \Delta Y_{\rm H^+} \varepsilon_{\rm H} \frac{k_{\rm H + \gamma \rightarrow H^+}} {\Sigma_i^n k_i +k_{\rm H + \gamma \rightarrow H^+} },
\ee
where $\Delta Y_{\rm H^+}$ is the difference of \hp is a time step, $\varepsilon_{\rm H}$ is the binding energy of hydrogen, $k_{\rm H + \gamma \rightarrow H^+}$ is the rate at which hydrogen is photoionized, 
%$k_{\rm H + \xi_H \rightarrow H^+}$ is the ionization rate from cosmic rays,   
and $\Sigma_i^n k_i$ is the summation of all the collisional ionization rates that form \hp . Similar equations can be written for other atomic species, such as helium and carbon, however, hydrogen always provides the dominant heating term. 

\subsubsection{Cooling Tests}

Having tested the chemistry package, we turned our attention to the cooling routines. Several of the cooling channels have been tested and used in previous applications, such as the high temperature atomic hydrogen and helium cooling and molecular hydrogen cooling (\eg GS10). For cooling from CO, OH, and \hto , we simply ensured that the interpolation between points in the table was smooth and returned the table values under the correct conditions. 

\section{Model Framework}

Once we had implemented the required chemistry and cooling routines into FLASH, we were then able to turn our attention to the structure and composition of the filament. We assume a Cold Dark Matter (CDM) cosmology with parameters, $\Omega_{\Lambda}$ = 0.7, $\Omega_{0}$ = 0.3, $\Omega_{b}$ = 0.045, and $h$ = 0.70, (\eg Spergel \etal 2007), where $h$ is the Hubble constant with units of 100 km s$^{-1}$ Mpc$^{-1}$, $\Omega_{\Lambda}$, $\Omega_{0}$, and $\Omega_{b}$ are the vacuum, total matter, and baryonic matter densities in units of the critical density. For our value of $h$, the critical density $\rho_{crit}$ is 9.30$\times$10$^{-30}$  gm cm$^{-3}$, which corresponds to a mean baryonic density of $\bar \rho_{b} = $4.18$\times$10$^{-31}$ gm cm$^{-3}$. 

\subsection{Filament Model}

Building on previous analytic and numerical results
we model the filament as a combination of dark and baryonic matter with a density profile (Stod$\acute{\rm o}$lkiewicz 1963; Ostriker 1964;  Harford \etal 2011) of
\be
\rho(\xi = r/r_s) = \frac{\rho_c}{(1+\frac{1}{8}\xi^2)^2} \ \ {\rm gm \ cm^{-3} },
\ee
where $\rho_c$ is the central gas density and $r_s$ is the scale radius.  If the filament were made of pure gas $r_s$ it would be in hydrostatic equillbrium if 
\be
r_s \equiv \frac{c_s}{(4 \pi G \rho_c)^{1/2}} = \left( \frac{k_b T}{4 \pi G \rho_c \mu m_p} \right)^{1/2} \ {\rm \ cm},
\label{rseqn}
\ee
where $c_s$ is the sound speed, $k_b$ is Boltzmann's constant, $\mu$ is the average molecular weight, $m_p$ is the mass of a proton, and $G$ is the gravitational constant. However, to account for the presence of dark matter we need to redefine this quantity, making use of the  
equation of hydrostatic equilibrium
\be
\nabla \Phi = \frac{1}{\rho_{\rm total}} \nabla P  = \frac{c_s^2}{(1+\beta) \rho_{g}}\nabla \rho_g,
\ee
where $P$ is the thermal gas pressure, $\rho_g$ is the baryonic gas density, and $\beta$ = $(\Omega_{0}- \Omega_{b})/\Omega_{b}$, where the constants are defined above. Using this equation we can also define an effective sound speed as
\be
c^2_{s,eff} \equiv c_s^2/(1+\beta).
\ee
Combining this with  Poisson's equation
\be
\nabla^2 \Phi = -4 \pi G \rho_{\rm total} = -4 \pi G (1+\beta) \rho_g
\ee
alters the definition of the scale radius to
\begin{eqnarray}
r_s &\equiv& \frac{c_s}{(4 \pi G (1 + \beta)^2 \rho_{g,c})^{1/2}}  \nonumber \\
       &=& \left( \frac{k_b T}{4 \pi G (1 + \beta)^2 \rho_{g,c} \mu m_p} \right)^{1/2}  {\rm \ cm} 
\end{eqnarray}
or
\be
 r_s= 68.3 \left[ \frac{T}{10^6 {\rm K}} \right]^{1/2} \left[ \frac{\Delta}{200} \right]^{-1/2} \left[ \frac{\mu}{0.6} \right]^{-1/2} {\rm kpc},
\ee
where we re-express the central density ($\rho_{g,c}$) as $\Delta \bar \rho_{\rm b}$ where $\Delta$ is a constant defined at runtime that represents the overdensity of the filament.

The composition of the filament is assumed to be initially ionized and atomic with abundances of $X_{\rm H}^+$ = 0.76, $X_{\rm He^+}$ = 0.24, $X_{\rm C^+}$ = 1.693$\times$10$^{-3}$Z$_{\rm frac}$, and $X_{\rm O^+}$ = 5.056$\times$10$^{-3}$ $Z_{\rm frac}$, where $Z_{\rm frac}$ is the metallicity of the gas in solar units and $X_{\rm i}$ is the mass fraction of the individual species. 

The long standing discrepancy between models of chemical evolution in galaxies and observations has given rise to the well known G Dwarf problem, first noticed by van den Bergh (1962) and Schmidt (1963). Simple  chemical evolution models produce a population of metal-poor stars that are not seen in observations. One solution is to assume that the gas has some initial metallicity, perhaps provided by outflows from dwarf galaxies at high redshift (\eg Scannapieco 2005). To that end, we assumed that in these low redshift cases that the filament gas has been enriched to $Z = 0.1 \, Z_{\odot}.  $

\subsection{Gravity and the Jeans Limit}

To account for the gravitational potential due to the surrounding dark matter, we calculate the gravitational acceleration of the gas as, $a_{\rm total} = a_{\rm SG} + a_{\rm DM}$, where $a_{\rm SG}$ is the self gravity of the gas as computed by the Poisson solver within FLASH (Ricker 2008), and the acceleration from the dark matter is given as
\be
a_{\rm DM} = -\nabla \Phi = - 2  c_s^2 \beta \, \nabla  {\rm ln}\left(1+\frac{1}{8} \xi^2 \right).
\ee
At the beginning of each simulation, the baryonic matter was assumed to be in hydrostatic equilibrium with the dark matter. A test of this gravity scheme showed that the filament remains in hydrostatic balance for many free-fall times in the absence of any chemistry or cooling. 

A common problem that arises in hydrodynamical simulations is artificial fragmentation of perturbations that comes from purely numerical sources. Truelove \etal (1997) studied this problem and found that to avoid this numerical artifact, the spatial resolution of the simulation must be smaller than the Jeans length, defined as (Jeans 1902),
\be
\lambda_j \equiv \frac{c_s}{\left( G \rho \right)^{1/2} },
\ee
where $c_s$ is the local sound speed. Truelove suggests that the ratio between the Jeans length and the spatial resolution, known as the Jeans number ($N_j = \lambda_j / \Delta x$), should be greater than 4 for hydrodynamical simulations, although, cosmological simulations have found that a value of 7 is required (Ceverino \etal 2010).  

With the advent of adaptive mesh refinement codes, this requirement becomes easier to manage since one can refine only those regions that require it. However, this is feasible only to a point. When the Jeans length gets below a few times the resolution of even the highest AMR level, we introduce artificial pressure support of the form,
\be
P_{\rm art} = C^2 G \rho^2 (\Delta x_{\rm min})^2 / \mu,
\ee
where $C =10,$ is a dimensionless constant proportional to the number of cells being smoothed over, $\rho$ is the density, and $\Delta x_{\rm min}$ is the minimum cell size (Machacek \etal 2001; Agertz \etal 2009; Ceverino \etal 2010; Klar \& M$\ddot{\rm u}$cket 2012) which is used in place of the thermal pressure in further hydrodynamic evolution. 

Naturally, once this condition is met, we can no longer follow the small scale density evolution of that region since the artificial pressure will act to smooth out these perturbations. Therefore, in the analysis that follows, we will describe these regions in terms of average quantities instead of specific values.  

Finally, the use of a pressure floor would cause the temperature to rise to unphysical values if not accounted for in the equation of state. To take this into account, we tracked the fraction of the pressure that comes from the artificial source and corrected the temperature for use in the chemistry and cooling routines. 

It should also be noted that we begin each simulation in hydrostatic equilibrium, but not in thermal equilibrium. The reason for this is two-fold. First, the cooling experienced from hydrogen-helium recombination and metal-line cooling is very strong, and second, the possible heating terms are very small.

The number of possible heating mechanisms is very limited and largely depend on the UV or cosmic ray background. In the absence of these conditions, the heating is negligible. Even with the presence of a nominal background, the photoelectric heating is only marginal with a rate of $\approx$ 1.0$\times$10$^{-26}$$n$ ergs s$^{-1}$ cm$^{-3}$, for a 10$^{6}$ K gas, an efficiency of $\approx$ 1, and $G = 0.01$, where $n$ is the number density.  Similarly for cosmic ray heating, the nominal value for the cosmic ray ionization rate is $\approx$ 1.0$\times$10$^{-17}$ s$^{-1}$ (Dalgarno 2006), which leads to a heating rate of $\approx 10^{-28} n$ ergs s$^{-1}$ cm$^{-3}$. Both of these rates are much smaller than the cooling mentioned above.

\section{Results}
%The first column is give the name of the run, the second and third columns give the initial gas overdensity and temperature respectively. The fourth column is the resulting scale radius of the cylinder in units of kpc. The fifth and sixth columns give the incident UV radiation field in units of the Habing field and the initial gas metallicity respectively. The final column gives the the Free-Fall time scale in units of Gyrs.{\tt ES Usually, one puts a bunch of notes to describe columns with unclear meanings rather than writing it all out like this.}
\begin{table*}
\caption{Summary of completed simulations. }
\begin{centering}
\begin{tabular}{lcccccc}
\hline
     Name & Overdensity & Temp. (K) & R$_{s} ({\rm kpc})$   & UV Background & Metallicity (Z$_{\sun}$)   & \tff (Gyrs)     \\
\hline
\hline
      Fiducial (FID) & 2000 & 1.0$\times$10$^6$ & 21.6   & 0.0 & 1.0$\times$10$^{-1}$  &  0.895   \\
      Ldens    &   200 & 1.0$\times$10$^6$ & 68.3   & 0.0 & 1.0$\times$10$^{-1}$  &  2.830   \\
      HightT  & 2000 & 1.0$\times$10$^7$ & 68.3   & 0.0 & 1.0$\times$10$^{-1}$   &  0.895   \\
      UV1     & 2000 & 1.0$\times$10$^6$ & 21.6   & 1.0 & 1.0$\times$10$^{-1}$    &  0.895   \\
      UV2     & 2000 & 1.0$\times$10$^6$ & 21.6   & 0.1 & 1.0$\times$10$^{-1}$    &  0.895   \\
      Highz   &   500 & 1.2$\times$10$^4$ & 0.83   & 0.0 & 1.0$\times$10$^{-3}$    &  0.385   \\
\label{simsum}
\end{tabular}  
\end{centering}
\end{table*}

Each of our simulations is carried out in two dimensions with cylindrical coordinates with the $r$ and $z$ axis spanning [0,5] \rs. If the ratio of the first and second derivatives of density, temperature, or pressure exceeded 0.6 divided by the cell size, then the block was marked for refinement. Conversely, if the value of this ratio for all three variables  was less than 0.20 over the cell size, it  was marked for derefinement. If a filament were to lose all thermal pressure support, it would collapse on a free-fall timescale given by
\be
\tau_{\rm ff} = \sqrt{  \frac{3 \pi}{32 G \beta  \Delta \rho_c} } \approx 0.9 \ \rm{Gyrs}
\label{tauff}
\ee
and, in addition to the criteria above, we forced derefinement in regions that were greater than 2.0 \rs \ from the center of the cylinder at times greater than 0.5 $\tau_{\rm ff}$. Each simulation  was run with a maximum of 11 refinement levels, which gave a large dynamic range of 8192 and a minimum cell size of  13.2 pc in our fiducial case. We imposed a reflecting boundary condition along the central axis and a diode boundary condition on the right, which allowed gas to leave the simulation area and prevents inflow from these boundaries.  We also impose diode boundary conditions at the top and bottom of the simulation domain.

A summary of the simulations performed is given in Table~\ref{simsum}. For each run we give the initial temperature, overdensity of the cylinder, corresponding scale radius, strength of the UV background, initial metallicity of the gas, and associated free-fall time.

\subsection{Hydrodynamic Evolution}
\label{fidevo}

\begin{figure*}
\centering
\includegraphics[trim= 0.0mm 0.0mm 0.0mm 0.0mm,clip,scale=0.40]{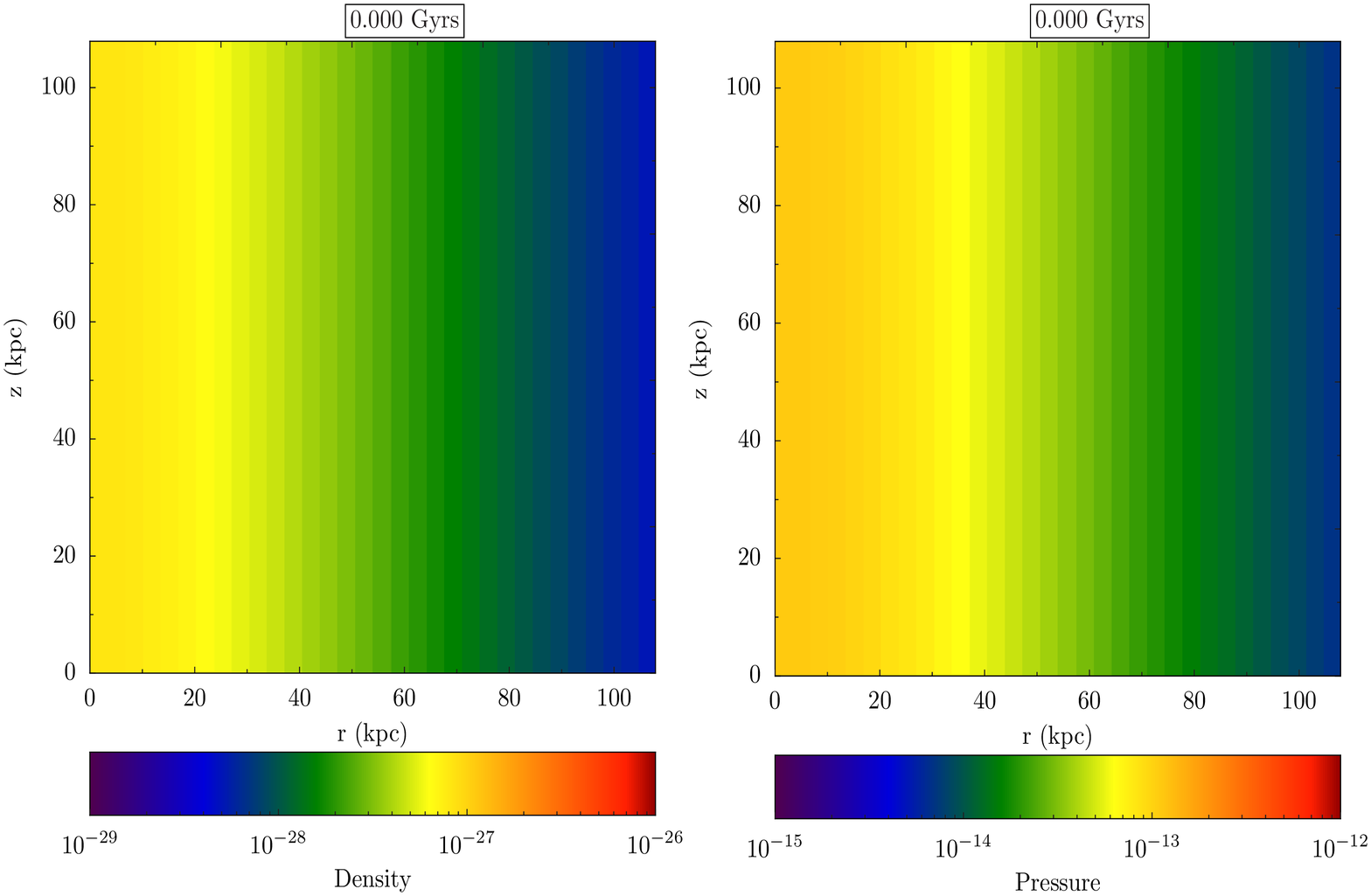}
\begin{centering}
\includegraphics[trim= 0.0mm 0.0mm 0.0mm 0.0mm,clip,scale=0.40]{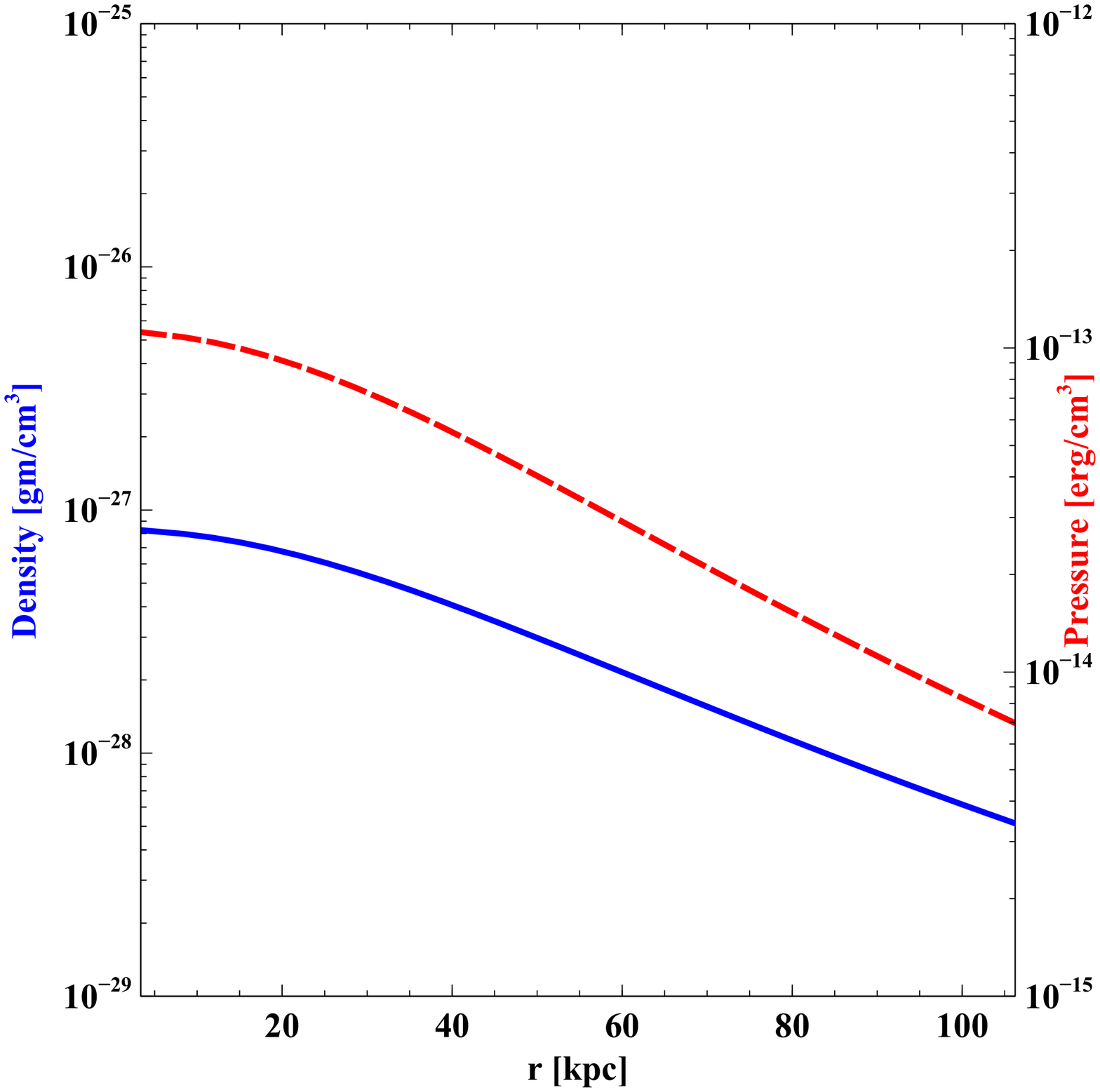}
\end{centering}
\caption{Initial simulation setup for our fiducial parameters. {\it Top Left}: Logarithmic density contours between 10$^{-28}<\rho< 10^{-25}$ g cm$^{-3}$. {\it Top Right:} Initial logarithmic pressure contours between 10$^{-15}< P < 10^{-12}$ ergs/cm$^{-3}$. {\it Bottom:} 1D profiles of the initial density (solid blue line) and pressure (dashed red line).  }
\label{init}
\end{figure*}

The initial density, pressure, and temperature profiles for our fiducial case are shown in Figure~\ref{init}. The cooling experienced by this gas largely determines its hydrodynamic evolution. Initially, cooling is dominated by atomic cooling from metals and hydrogen and helium.
The atomic cooling time scale, \ie\ the time required to completely cool the gas from an initial temperature, $T,$ is given by
\be
\tau_{cool} = \frac{1.5 n k_b T}{ n_{H} n_{e} \left(\Lambda_H + Z_{\rm frac} \Lambda_M  \right) } \ {\rm (s) } ,
\label{taucc}
\ee
where $n$, $n_H$, and $n_e$ are the total number density, hydrogen number density, and electron number density respectively,  and $\Lambda_{\rm H}$ and $\Lambda_{\rm M}$ are the atomic and metal-line cooling rate respectively. For our initial temperature of 10$^6$ K, this time scale is $\approx 1.8$ Gyrs. However, cooling becomes more efficient at slightly lower temperatures ( $\sim 10^5$K, \eg Sutherland \& Dopita, 1993) until the hydrogen-helium cooling floor is reached at 10$^4$ K, at which point the timescale is a fraction of a Gyr.

\begin{figure*}
\centering
\includegraphics[trim= 0.0mm 0.0mm 0.0mm 0.0mm,clip,scale=0.75]{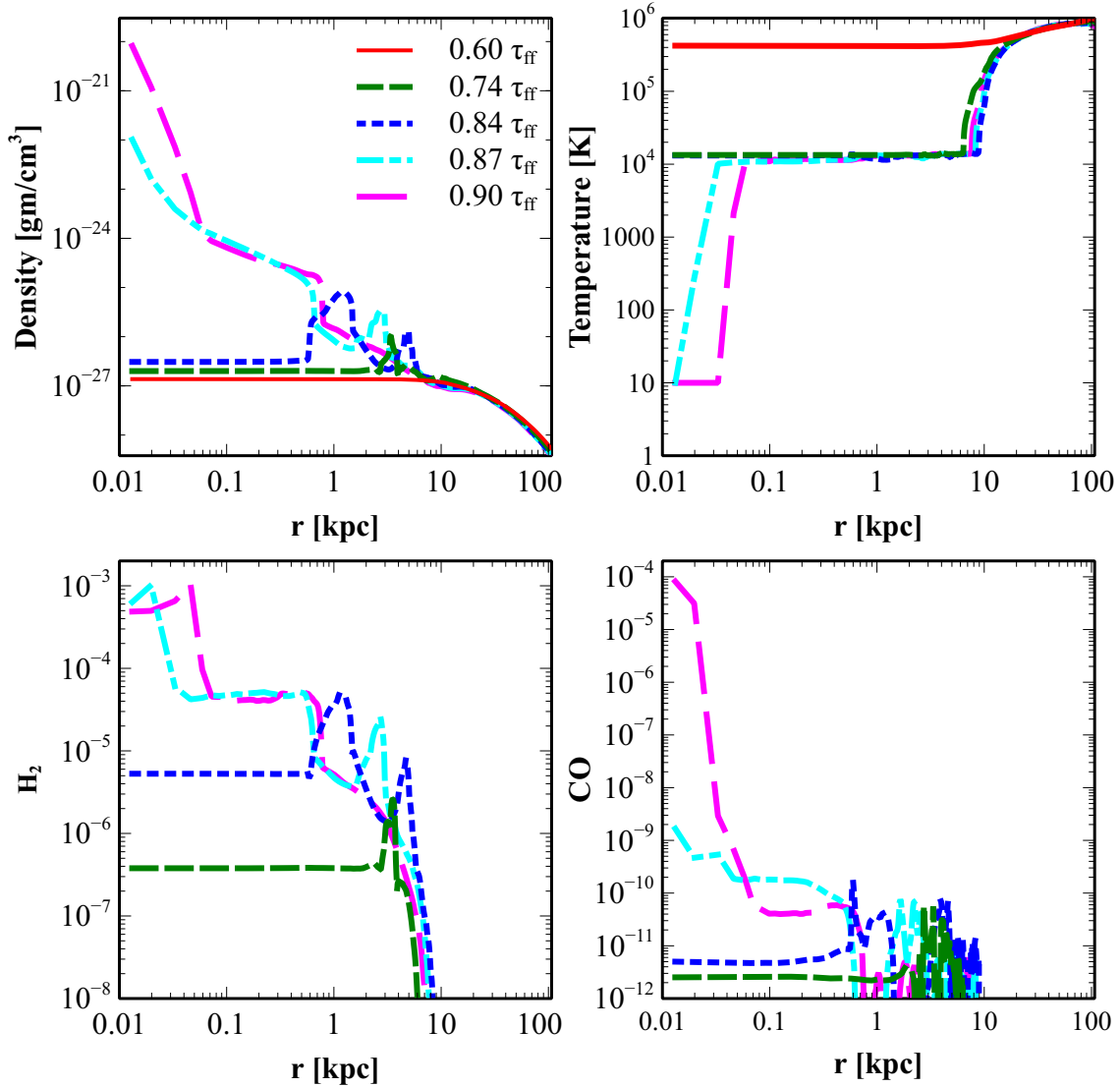}
\caption{1D profiles from the fiducial run. The $x$-axis in each panel is the logarithm of the radial position along the filament. Each $y$-axis is the logarithm of each variable. The top left panel shows the density profiles, the top right the temperature profiles, bottom left shows the H$_2$ mass fraction, and the bottom right shows the CO mass fraction. Times are given in the legend in the density plot in units of the initial free fall time. }
\label{evo1}
\end{figure*}

\begin{figure*}
\centering
\includegraphics[trim= 0.0mm 0.0mm 0.0mm 0.0mm,clip,scale=0.80]{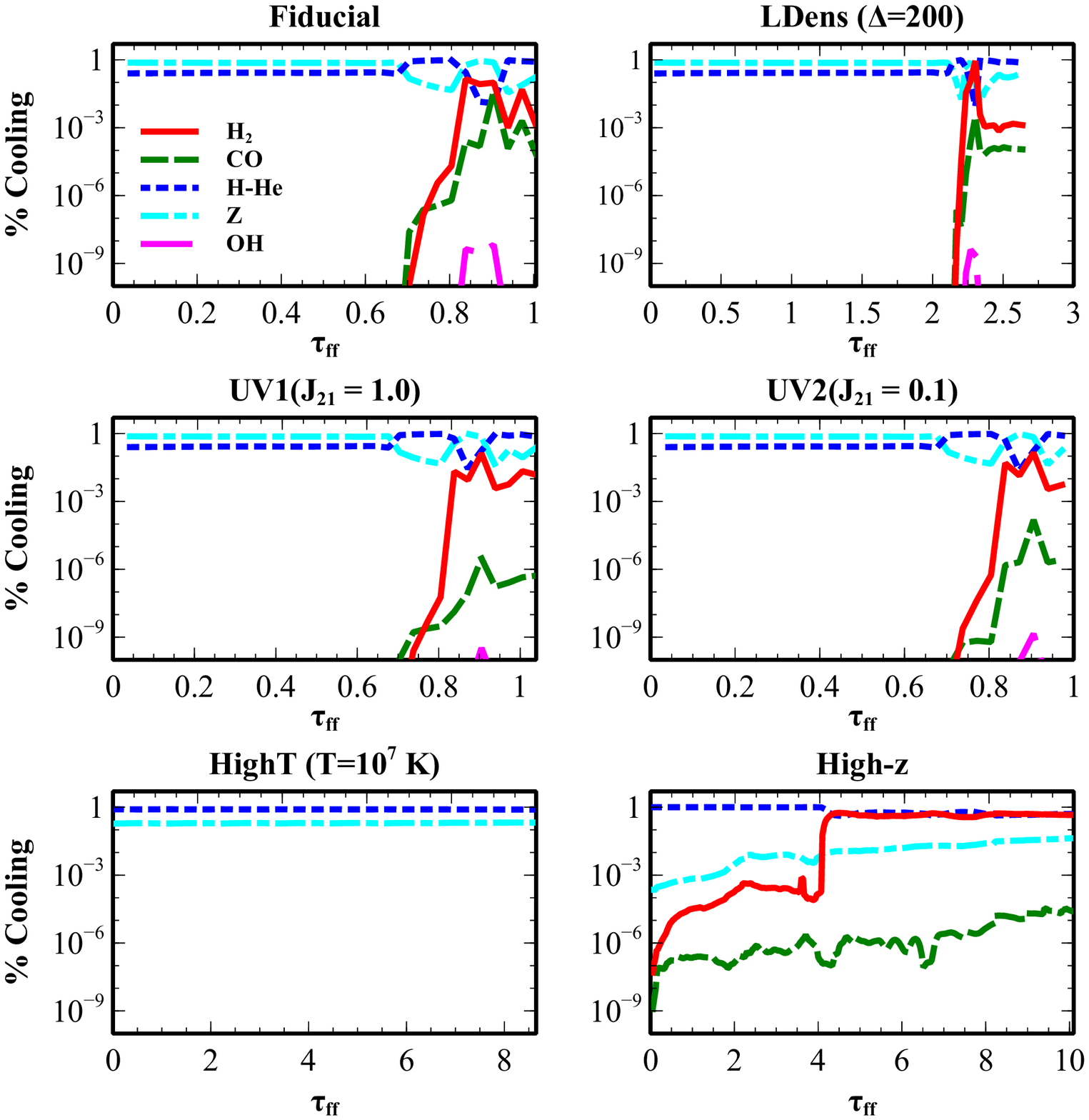}
\caption{Summary of the density averaged cooling for each simulation. The top left panel shows the cooling terms for our fiducial run, the top right shows Ldens, the middle left is for UV1, the middle right shows UV2, the bottom left is for HighT, and the bottom right shows Highz. Each panel shows the density weighted cooling for each of the primary cooling terms. The cooling ratio is given on the $y$-axis and is calculated by dividing each cooling term by the sum total of all cooling terms. The simulation time is given of the $x$-axis in the units of the free fall time. The blue (dotted) line shows the contribution from atomic hydrogen-helium cooling, the cyan (dash-dotted) line depicts the metal line cooling, the red (solid) lines is cooling from molecular hydrogen H$_2$, the green (dashed) is from CO, and the magenta (fine dashed) line is cooling from OH. Note that although we also track the heating from molecular hydrogen dissociation from the UV background and cooling from water (H$_2$O) neither one of these processes is efficient enough to be plotted in these panels.  }
\label{cool1}
\end{figure*}

In Figure.~\ref{evo1}, we show 1D profiles of the evolution of our fiducial filament. The top left panel shows the density profile, the top right shows the temperature profile, the bottom left shows the H$_2$ mass fraction, and the bottom right panel shows the CO mass fraction.  As the filament gas begins to cool, the initially ionized gas rapidly recombines and cools to near the hydrogen-helium cooling floor. At this point, the gas begins to collapse toward the center of the filament which increases the density, and therefore, the cooling experienced at the center. The relative importance of each cooling term is shown in the top left panel of Figure~\ref{cool1}. Early on, cooling is dominated by atomic hydrogen-helium and metal-line cooling. However, once the cloud becomes dense enough for molecules to form, their cooling channels become as important as the atomic cooling. 

As the gas cools, shocks are formed as the infalling gas interacts with this cooler interior gas. These create peaks in the density profiles and allow for an increase in the abundances in molecular species (cyan and magenta lines in Fig~\ref{evo1}). The shocks are then reflected off the central axis.  However, these density peaks are quickly erased and the gas returns to a much more uniform distribution as shown by the dash-dotted and long-dashed lines in Figure~\ref{evo1}. 

The final state of the filament is shown by the long-dashed  lines in Figure.~\ref{evo1}. The gas has cooled and collapsed into a  compact structure with a high central density ($\rho \approx 1.0\times10^{-21}$ gm cm$^{-3}$) and cold central temperature ($T \approx 10$ K). In addition, the gas is enriched with molecular coolants with interior abundances of 10$^{-3}$ for H$_2$ and 10$^{-6}$ for CO respectively. These values should be taken as averages since at these temperatures we are forced to use our alternative pressure scheme to prevent unphysical collapse. 

\subsection{Effect of Density}

The variable $\Delta$ controls the many of the properties of the filament, including its central density, size, and evolution timescale. More importantly, both chemistry and cooling rates are dependent on this density but with different scalings.  As can be seen from eq.~(\ref{tauff}), the free-fall timescale scales as $\rho^{-1/2}$ while the cooling timescale (eq.~\ref{taucc}) goes as $\rho^{-1}$.  Thus, in addition to our fiducial run, we ran a simulation where $\Delta$ was reduced by a factor of 10.

\begin{figure*}
\centering
\includegraphics[trim=0.0mm 0.0mm 0.0mm 0.0mm,clip,scale=0.7]{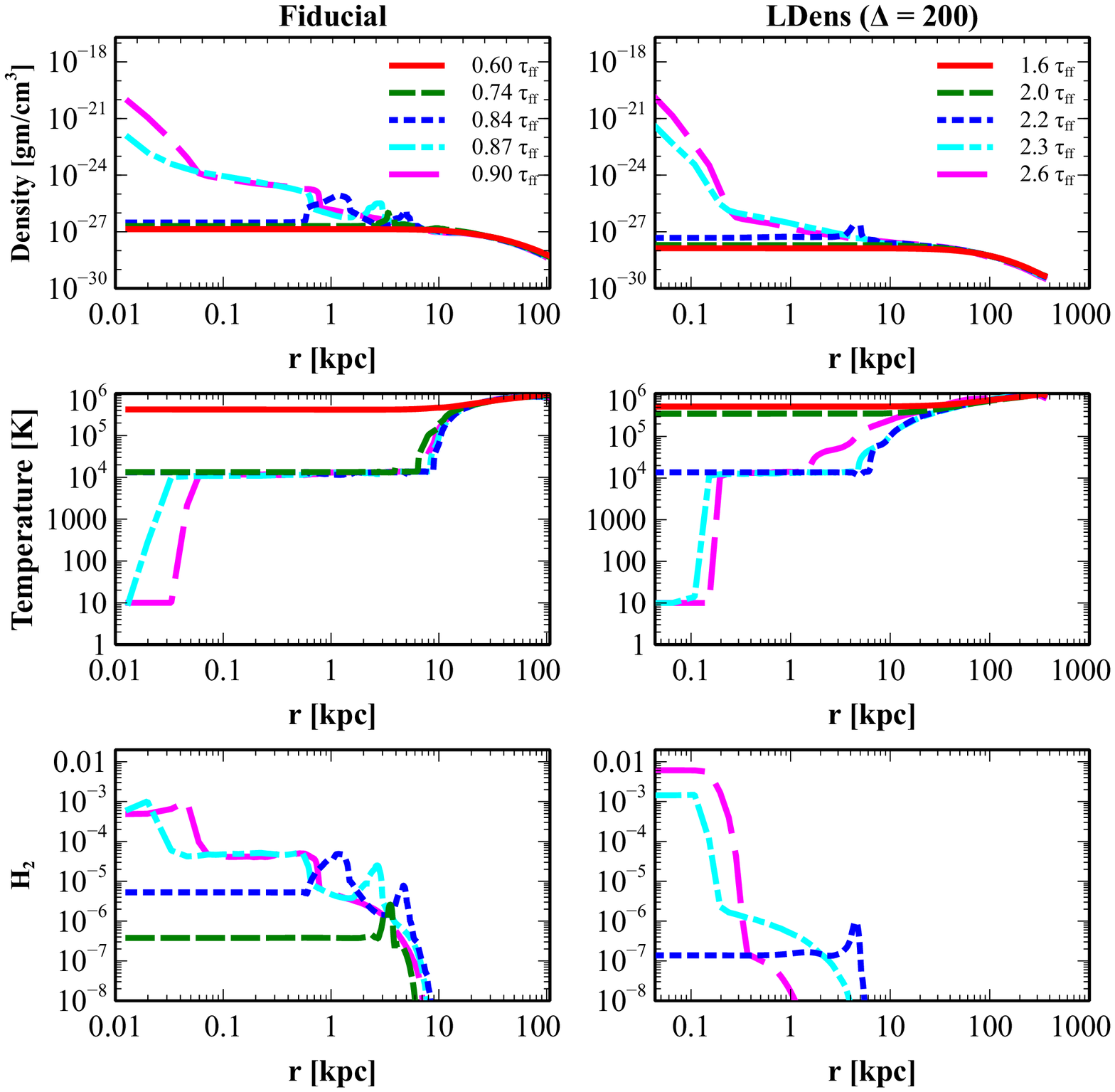}
\caption{Comparison between the fiducial and low density cases. The top row shows the 1D density profiles at similar evolutionary stages, the middle row shows the temperature, and the bottom row shows the molecular hydrogen (H$_2$) mass fraction. The first column shows the fiducial case while the second column shows the low density case. The $x$-axis is the logarithm of the radial position along the filament. Note that in the low density case there are fewer density peaks due to shocks. However, the final state of the filament is similar.}
\label{ldcomp}
\end{figure*}

Figure~\ref{ldcomp} compares the evolution of  the density, temperature, and H$_2$ mass fraction in this simulation with the fiducial run at similar evolutionary stages. In general, the low density and fiducial cases evolve in similar ways. Initially, the gas recombines and the filament begins to collapse, but the cooling in the lower density gas is not as efficient. This prevents shocks from forming and allows for a much more uniform collapse. Ultimately the outcome is the same, however, a dense core is formed at the center of the filament,  with a large molecular fraction.

The top right panel of  Figure~\ref{cool1} shows the relative importance of each cooling term. Again, there is a strong similarity with the fiducial case. In each case, atomic hydrogen-helium and metal-line cooling dominates early on and, after an abundance of molecular coolants are formed, they become an important cooling channel. An important note is that while in the fiducial case the whole filament collapses within a single free-fall time (0.905 $\tau_{\rm ff}$ in the fiducial case), the low density case takes much longer (2.63 $\tau_{\rm ff}$), due to the increase in the cooling time, which scales as one over the number density. This can been seen when comparing the results in the top row of Fig~\ref{cool1}. In the fiducial case, for example, molecular cooling becomes important at $\approx$ 0.7 $\tau_{\rm ff,}$ while in the low density case, it does not become important until $\approx$ 2.3 \tff.

\subsection{Effect of UV Background}

Next we explored the impact of a dissociating UV background on the chemical and hydrodynamical evolution of the filament. We ran two simulations with different UV backgrounds, one with a full strength Habig field (UV1), denoted by a J$_{21}$ value of 1.0 and a field with one tenth that strength (UV2). This field has the primary effect of dissociating molecular coolants after they have formed, in particular, molecular hydrogen (H$_2$) and carbon monoxide (CO). 

\begin{figure*}
\centering
\includegraphics[trim=0.0mm 0.0mm 0.0mm 0.0mm,clip,scale=0.5]{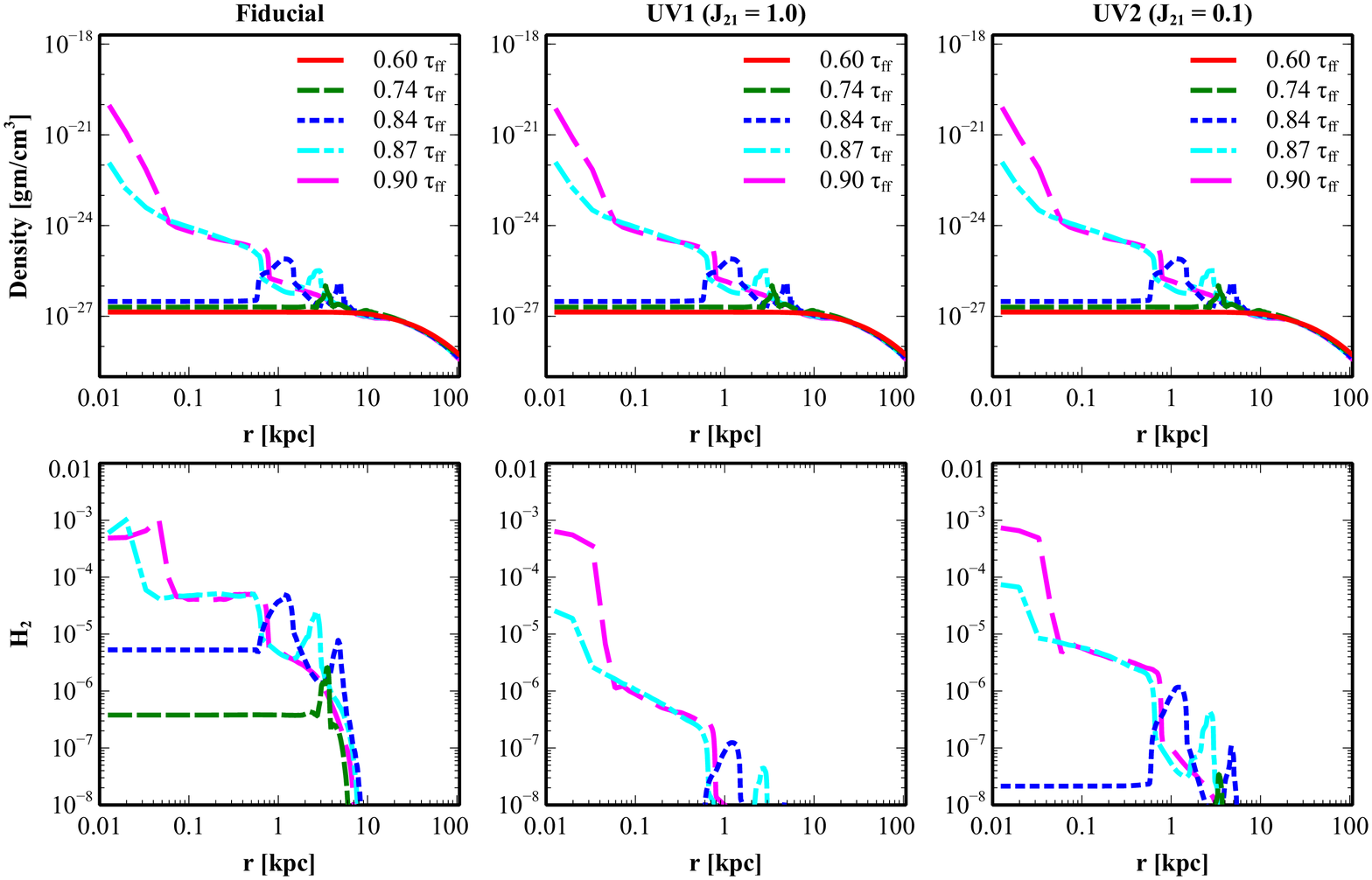}
\caption{Comparison of varying background UV strength. The first column shows the fiducial case (FID), the middle column shows UV1, and the last column shows UV2. The top row shows the 1D density profiles and the bottom row shows the molecular hydrogen mass fraction profiles. As the UV background increases, the more the formation of molecular coolants are suppressed. However, the final state of the filament changes little between each of these run. }
\label{uvcomp}
\end{figure*}

Figure~\ref{uvcomp} shows the 1D profile comparisons between these runs. The first column shows our fiducial case, the second column shows UV1, and the third column shows UV2. Since the free fall time for these simulations is the same, each of the lines corresponds to the same evolutionary time. Comparing the density profiles between these run (first row of Figure~\ref{uvcomp}) shows that there is little difference. All three evolve with the same shock structure developing and with similar final central densities. Slight differences can be seen in the temperature profiles where the fiducial case cools slightly faster than either UV1 or UV2 at late times. As expected, the greatest differences can be seen in the molecular hydrogen profiles. The stronger the UV background the lower the abundance of molecules. However, this has little impact of the final state of the filament, the cooling experienced, or even the final abundances found at the center of the filament.

The middle row of Figure~\ref{cool1} shows the relative importance of the various coolants for these runs. The primary difference is that while cooling from CO is  important for FID, it is much less important for the UV runs. Cooling from OH, while marginal for FID is non-existent in UV1 and UV2. The UV background also effects the time at which molecular cooling becomes important. The lower the incident background, the earlier the cooling is important. 

\subsection{Effect of the Initial Temperature}

The initial temperature of the filament not only determines the physical scale of the filament (eq.~\ref{rseqn}), but also the cooling rate experienced by the gas. To study the impact that the initial temperature has on the evolution of the filament, we run a simulation with an initial temperature of 10$^{7}$ K (HighT). 

In fact, it is the cooling rate that completely determines the fate of this filament. With the initial temperature so high, the cooling timescale becomes incredibly long. In fact, with an initial central number density of $\sim$10$^{-4}$cm$^{-3}$ and an initial temperature of 10$^{7}$ K, the cooling timescale is greater than 10 Gyrs, whereas the free-fall timescale is only 0.9 Gyrs. This allows the filament to remain close to hydrostatic balance over several filament free fall times.

\subsection{High Redshift Filament Evolution}

Finally, we explored the evolution of a high-redshift filament. Recent simulations by Harford \etal (2011) studied cosmological simulations at a $z=5$ and found baryon-rich filaments that stretched between galaxies. They found that these filaments were well described by an isothermal cylinder with a typical overdensity above the cosmic mean density of 500, temperatures of $\approx$ 1-2$\times$10$^{4}$ K, and nearly fully ionized gas. We use these values as initial conditions, also assuming a lower metallicity of $Z/Z_{\odot}=10^{-3}$, in this case,  since by this age in cosmic time, fewer stars have formed and have polluted the surrounding gas.  

\begin{figure*}
\centering
\includegraphics[trim=0.0mm 0.0mm 0.0mm 0.0mm,clip,scale=0.80]{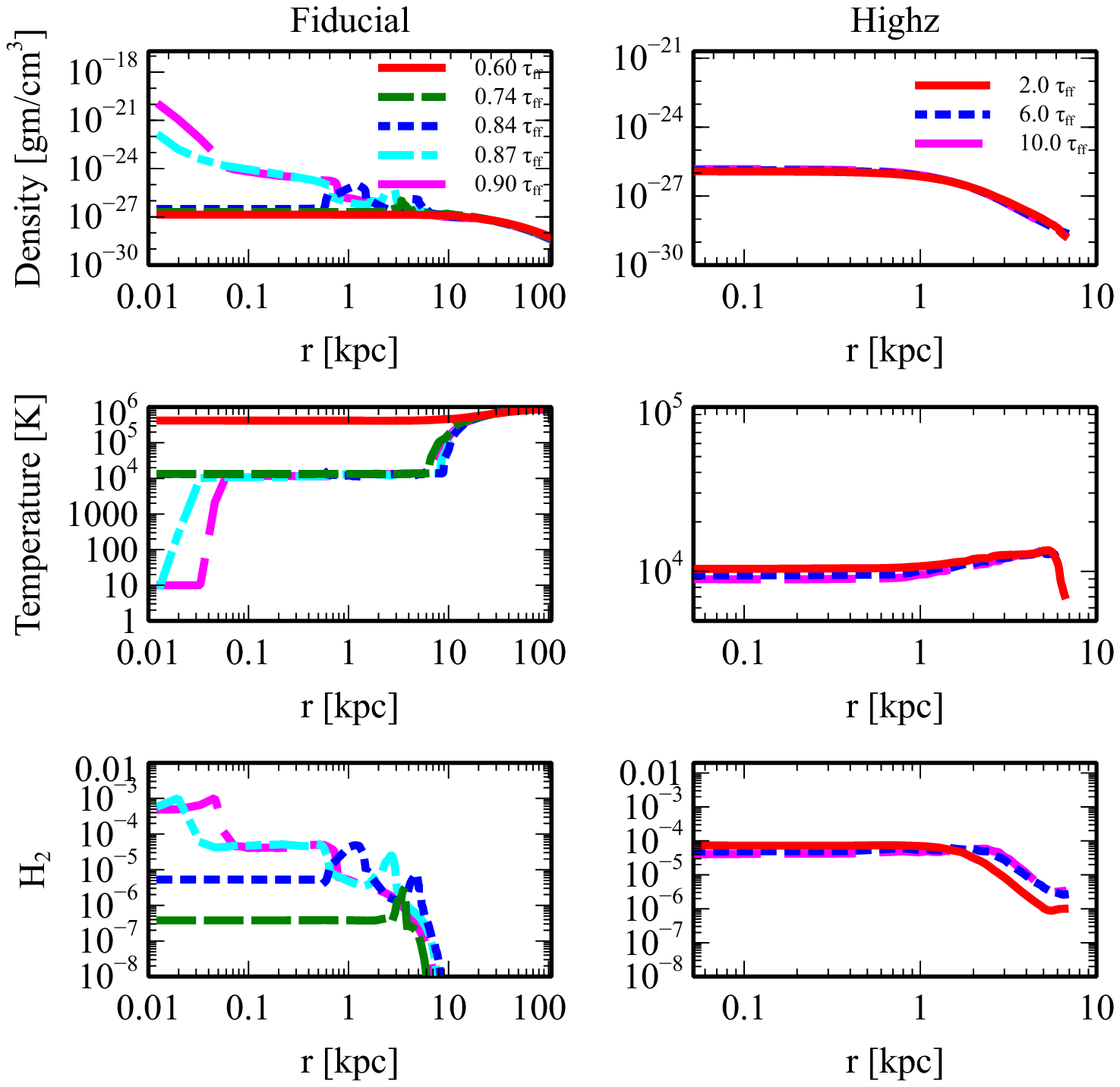}
\caption{Effect of redshift on filament collapse. The first column show FID while the second column shows Highz. Each row has the same meaning as Fig.~\ref{uvcomp}. Similar to the HighT case, the filament does not collapse but relaxes into a hydrostatic case. The cooling experienced in the high-redshift case differs from FID in that molecular hydrogen is more important than metal-line cooling. }
\label{highz}
\end{figure*}

Figure~\ref{highz} shows the comparison between Highz and FID. As in the high temperature case, the filament does not collapse and instead relaxes into a hydrostatically balanced state. That initially ionized gas, again, quickly recombines which allows for the formation of a small abundance of molecular coolants. However, since the initial temperature is close to the atomic hydrogen-helium cooling limit, the increase in density of the filament during the atomic cooling stage is minimal.  This means that densities are too low for molecular hydrogen, which is a much less efficient coolant, to take over at this stage and continue the collapse of the filament.  Instead, the filament reaches a new quasi-steady state, which it will maintain for many dynamical times.

The bottom right panel of Figure~\ref{cool1} the relative cooling experienced. This differs from FID in two aspects. First, since the metallicity is much lower in Highz than in FID, the relative contribution of metals is much lower. Second, since the Highz filament is initially much denser and cooler, molecular cooling is more important much earlier. Finally, over much of the evolution of the filament, cooling from H$_2$ is more important than metal-line cooling. However, this cooling is still not enough to allow the filament to continue to continue its collapse after the initial atomic cooling stage.

\section{A Filament Near a Dwarf Galaxy}

Dwarf galaxies are unique places in the study of star formation. Without the aid of normal dynamic causes, such as galactic bars or spiral arms (\eg Vanzi \etal 2009), dwarf galaxies experience the same level of star formation per unit mass as their larger cousins. This intense, episodic star formation might be caused by several possible mechanisms, including mergers, tidal interactions, and accretion from surrounding filaments (\eg Cumming \etal 2008; James \etal 2010). Meier \etal (2002) also suggest that due to the low metallicity of these galaxies, they represent an unique window into the early epochs of start formation.

NGC 5253, a nearby dwarf galaxy at a distance of 3.8 Mpc (Gibson \etal 2000) and companion to M83, is an excellent example of such a galaxy. Recent observations of this galaxy have uncovered intense bursts of star formation. Multiple ``super-star clusters" are extremely bright in the infrared with luminosities of $L_{IR}\sim  10^{9} L_{\odot}$ (\eg Gorijian, Turner, \& Beck 2001) and visual magnitudes of $M_V \sim$ -10 to -11 (Meurer \etal 1995; Gorjian 1996). Finally, dust lanes with motion consistent with inflow have been uncovered in radio observations of CO (2-1) (Meier \etal 2002). However, how much the inflow effects the nature of the starburst remains an open question.

The gravitational potential in the previous simulations have all been solely aligned with the radial direction and the resulting collapse is fairly uniform. However, in a physical situation, this is not the case. These filaments also feel the gravitational acceleration from nearby galaxies onto which they are accreted. The question we aim to answer is whether or not this additional acceleration is the mechanism by which the core of the filament can fragment and form dense clouds full of molecules.

\begin{figure*}
\centering
\includegraphics[trim=00.0mm 0.0mm 0.0mm 0.0mm,clip,scale=0.6]{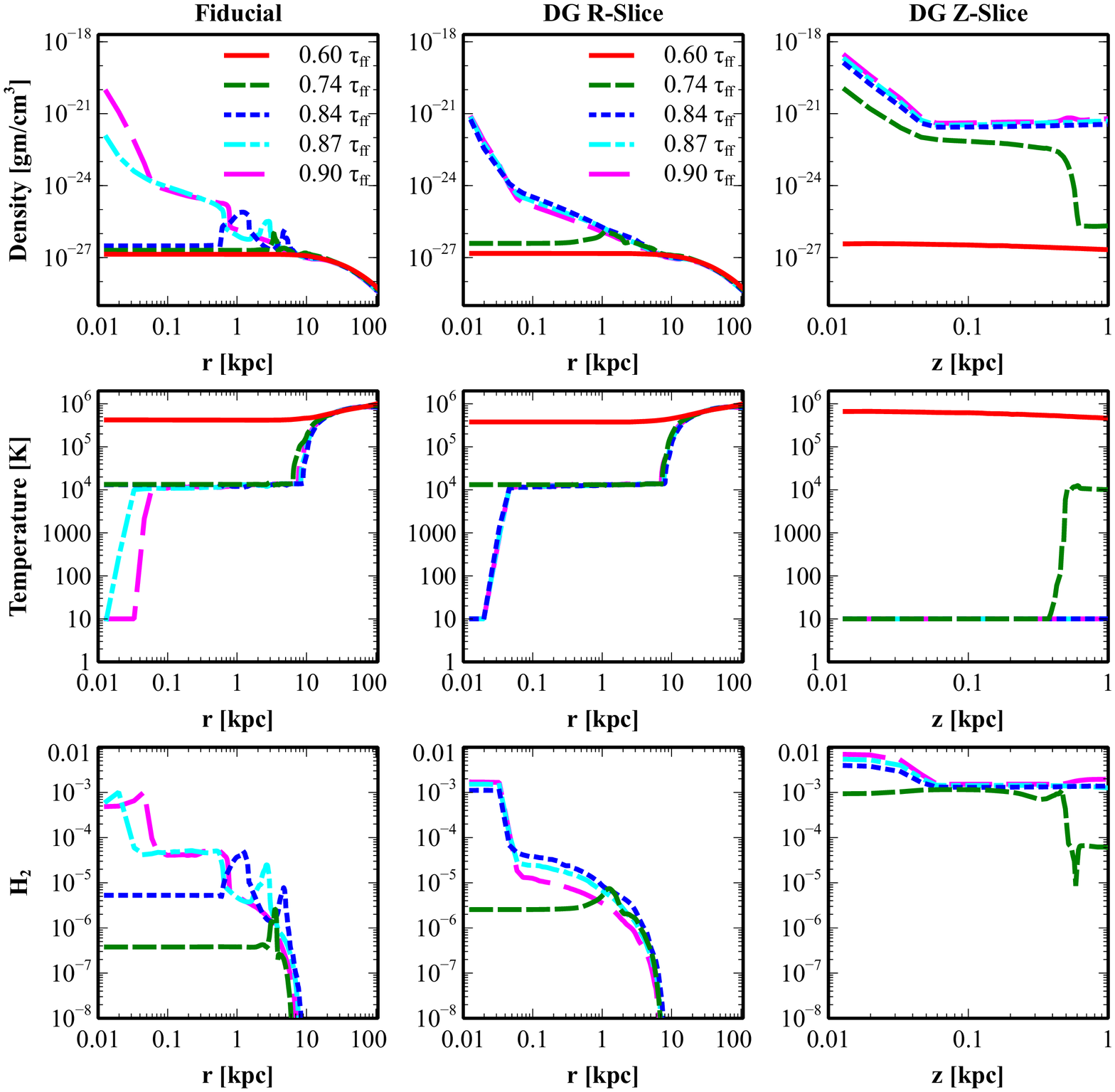}
\caption{Comparison between FID and a run with an additional gravitational potential due to a nearby dwarf galaxy. The first column shows the 1D profiles along the radial direction from FID, the second column shows the 1D profiles along the $r$-axis through the center of the simulation 54 kpc away from the center of the dwarf halo, and the third column shows the 1D $z$-axis profiles along the left $r$-axis boundary. The top row shows the density profiles, the middle row the temperature profiles, and the bottom row shows the H$_2$ mass fraction profiles. Again, we take the logarithm along the radial direction to highlight the profiles near the center of the filament (Columns 1 and 2) or in the center of the dwarf galaxy potential (Column 3).}
\label{dggcomp}
\end{figure*}

To this end, we have implemented an additional gravitational source term to account for the galaxy. The galaxy potential is modeled as an NFW halo (Navarro, Frenk, \& White 1997) which has a total interior mass as a function of radius of 
\be
M(<R) = 
\begin{cases} 
M_{\rm total} \, \frac{F(c R/R_{\rm max})}{F(c)} & \text{ if $R \leq R_{\rm max},$} \\
M_{\rm total}  & \text{if $R \ge R_{\rm max},$}
\end{cases}
\ee
where $M_{\rm total}$ is the total mass of the halo, $F(t)\equiv$ ln$(1+t) - \frac{t}{1+t},$ $c$ is the concentration parameter which we take to be 13.5 (Del Popolo \& Cardone 2012),  and $R_{\rm max} = 3 M_9^{1/3} \left( \frac{1+z_c}{10} \right)^{-1} \ {\rm kpc}$ is the virial radius of the galaxy halo, with $M_9 \equiv M_{\rm total}/ 10^9 M_{\sun}.$
This is given as given as 1.14$\times$10$^{3}$ (L$\acute{\rm o}$pez-S$\acute{\rm a}$nchez \etal 2012), where we take $z_c$(= 9) to be the virialization redshift.  
The gravitational acceleration from this profile is then given simply as  
\be
a_{\rm dg,grav} = -\frac{G M(<R)}{R^2}.
\ee

Figure~\ref{dggcomp} shows the 1D line profiles for density, \htwo, and \co\ for both FID and a run with the dwarf galaxy gravitational potential, where profiles in both the $r$-direction (through the center of the simulation, 54.0 kpc away from the center of the dwarf halo) and the $z$-direction (along left $r$-axis). The evolution of the filament remains largely the same as the fiducial case. The filament gas recombines and collapses toward the center of the dark matter potential. In both cases the center of the filament becomes very dense and very cold.

An interesting difference between these two cases, however, is the strength of the shock formed as the the filament collapses. The strength of the shock is related to how quickly the filament collapses. The evolution of the fiducial case is outlined in \S \ref{fidevo}. On the other hand, the run with the dwarf galaxy the evolution is slightly faster. This leads to an increase in the density near the center of the filament as compared to the fiducial case, which causes the shocks to be less pronounced. However, these shocks are largely transient and don't impact the final state of the filament in a large way. 

% In addition, the potential from the dwarf galaxy pulls much of the filament gas toward the center of that potential. This is shown in the third column of Figure~\ref{dggcomp}. By the end of the simulation, the density at the center of the dwarf galaxy is nearly an order of magnitude higher than the density at the center of the fiducial filament. 

%The gas at the center of the dwarf galaxy evolves at a quicker rate than compared to other regions of the filament. This is simply attributed to the increase in density and, therefore, the increase in number density which increases the rate at which chemical species are created and the quicker the gas will cool. This is shown by comparing the second and third column of Figure~\ref{dggcomp}. In addition, the molecular gas fraction is nearly an order of magnitude higher here than in the rest of the filament. 

\begin{figure*}
\centering
\includegraphics[trim=0.0mm 0.0mm 0.0mm 0.0mm,clip,scale=0.50]{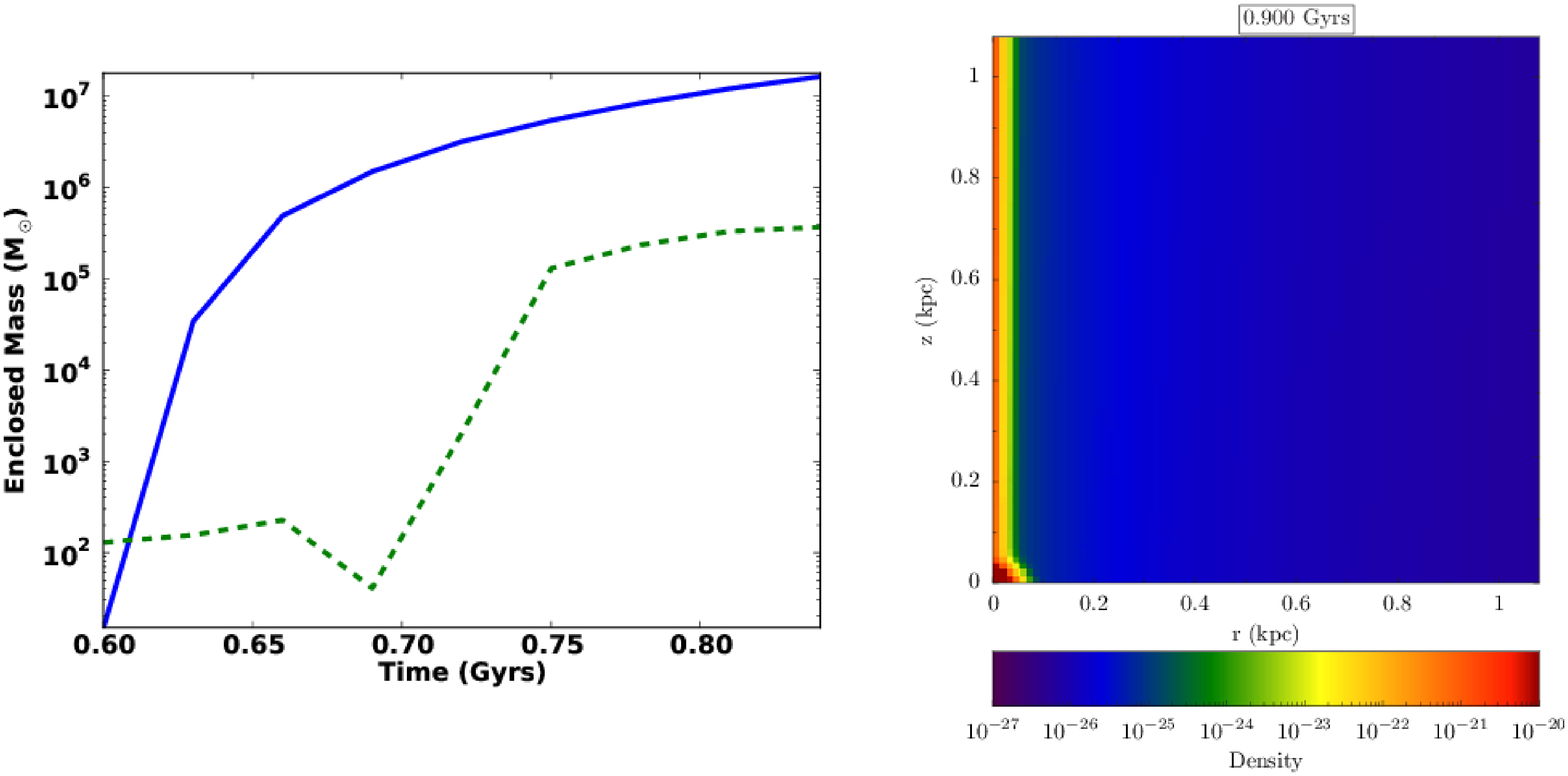}
\caption{ {\it Left:} Comparison between two equally sized regions, one containing the center of the dwarf galaxy and the other centered on the collapsing filament. As expected the region with the extra potential contains much more mass with an accretion rate of $\sim$0.1 M$_{\odot}$/yr. {\it Right:} A 2D logarithmic density contours at the end of dwarf galaxy simulation showing the density peak at the center of the dwarf galaxy potential.}
\label{dggsum}
\end{figure*}

The left panel of Figure~\ref{dggsum} shows a 2D logarithmic density profile where the center of the galaxy potential is at [0,0] at the end of the simulation. The profile shows an ideal situation for the continued accretion of gas into the dwarf galaxy. The density peaks at the center of the dwarf galaxy potential while the rest of the filament gas has collapsed into a dense, cold structure stretching along the interior $r$-axis. Once the gas has collapsed into the center of the dark matter potential, it is susceptible to collapse into the dwarf galaxy. This influx of cold dense gas promotes further star formation within the galaxy. 

Finally, the right panel of Figure~\ref{dggsum} shows the comparison of the accreted mass in two equally sized regions in the filament. The solid (blue) line shows a [108,108] pc region centered on [54,54] pc and includes the region nearest the center of the dwarf galaxy. The dashed (green) line shows a comparable region centered on [54 pc, 107.8 kpc] which is likely to be the least altered by the gravity from the dwarf galaxy. We plot the evolution of the enclosed mass these regions versus time once the filament starts to collapse. As expected, the addition of the dwarf galaxy drastically increases the amount of gas collected in that region. From this, we can estimate an accretion rate of $\sim 0.1$ M$_{\odot}$/yr which, if most of this gas is turned to stars, matches the star formation estimate given by L$\acute{\rm o}$pez-S$\acute{\rm a}$nchez \etal (2012) ($\sim 0.2-0.1$ M$_{\odot}$/yr).  

Many observations of NGC 5253 have uncovered multiple giant molecular clouds (GMC) that are found in and around the central starburst and are thought to be the product of infalling gas (Meier \etal 2002). The formation of these GMCs are an open question, although some have suggested that they were formed from an interaction between NGC 5253 and M83 (\eg van den Berge 1980; Kobulnicky \& Skillman 1995; Calzetti \etal 1999). Although most of the GMCs are not resolved in their radio observations, Meier \etal (2002) reports estimated physical sizes of these clusters to be $<$ 89 pc in diameter. Studies of other, less active, dwarf galaxies have GMCs with sizes between $\sim$ 10-70 pc (\eg Rubio, Lequeux, \& Boulanger 1993; Wilson 1994; Wilson 1995; Taylor \etal 1999; Walter \etal 2001; Meier, Turner, \& Beck 2001). To determine whether or not this occurs in our simulations, we reran the dwarf galaxy simulation with an additional two levels of refinement. This gives us a resolution of 3.5 pc at the finest level. The evolution of the filament remains largely the same save the higher central density (by about an order of magnitude) at the center of filament at the end of the simulation.  

Unfortunately, such clusters are not formed in any of the simulations presented.  On the other hand, it is still unclear to what extent their absence is due to the axisymmetric geometry imposed, which may underestimate fragmentation due to cooling instabilities as the filament collapses. Furthermore, due the small sizes of some GMCs, even in the extremely-high 3.5 pc resolution case, it is possible that we are not resolving the formation of these GMCs due in part by the limit on refinement level and use of the artificial pressure.  Future, extremely high-resolution, fully three-dimensional simulations will be necessary to settle this issue definitively.
 
\section{Discussion \& Conclusion}

Filaments are ubiquitous structures that stretch between the galaxies and dark matter halos found in observations and large-scale N-body cosmological simulations. In galaxies like the Milky Way, they provide the gas reservoir to continue the observed near constant star formation. In smaller galaxies, however, the effect is not quite as peaceful. Dwarf galaxies experience episodes of star formation that are just as strong as those seen in larger galaxies. In addition, dwarf galaxies are more metal-poor than their larger counterparts, which provides a unique window into the process of star formation at earlier star formation epochs. However, the mechanism that causes such extreme star formation in these small galaxies is not well understood, although mergers and infalling gas may be viable candidates.

In this paper we have studied how these initially ionized, hydrostatic filaments evolve in time by tracking the formation and evolution of key molecular coolants and their associated cooling. In the fiducial case, much of the early evolution is determined by atomic cooling from metals and hydrogen-helium lines. As soon as the filament gets dense enough important molecules are formed which then allows the filament gas to cool to below the hydrogen-helium floor. The initial cloud has collapsed into the center of the original filament and is characterized by having a very dense and cold core. This transformation takes nearly a single free-fall time to occur.

The inclusion of a pressure floor to preserve the Jeans criterions prevents us from following the specific details of the filament once the gas becomes very cold and dense. With improved resolution, the collapse could be continued down to much higher densities,  to determine in better detail if the conditions and scales are reached at which stars are expected to form. However, pushing to such high densities is beyond the scope of this paper. 

The inclusion of a dissociating UV background has little impact on the final state of the filament. As expected the inclusion of the background impacts the time at which molecular cooling becomes efficient as some molecules are destroyed. However, the final state of the filament remains the same as the fiducial case. 

This story is not repeated for other initial conditions. In both HighT and Highz the filament does not evolve drastically from its initial profile. This is due to the extremely long cooling time scales encountered during the collapse. In the HighT case, the low density and high temperatures of the filament results in a cooling time scale that can be up to several orders of magnitude greater than the free-fall time scale. Highz begins near the atomic hydrogen-helium cooling floor, and thus does not increase its density sufficiently during the atomic cooling stage for molecular coolants to efficiently continue the collapse.

Finally, we studied the effect that the gravitational potential from a nearby dwarf galaxy has on the collapse of the filament. While much of the evolution remains the same as the fiducial case, a density peak forms at the center of the galaxy potential. Due to the increase in density, and hence number densities, this region undergo quicker chemical and thermal evolution than the rest of the filament. 

We also compared this situation to NGC 5253, a starbursting galaxy located nearby at a distance of 3.8 Mpc. The starburst is thought to be powered by infalling filamentary gas similar to the situation studied, and in fact, our simulations reproduce a star formation rate of $\sim 0.1$ M$_{\odot}$/yr if we assume all of the infalling gas is converted to stars, which is similar to the estimate given by L$\acute{\rm o}$pez-S$\acute{\rm a}$nchez \etal (2012).  On the other hand, the multiple GMCs seen in radio observations are not reproduced in our simulations. This may be due to the need for full three-dimensional simulations and perhaps even higher resolutions to fully capture the fragmentation process during the collapse of the filament and its accretion onto the dwarf galaxy.   For now, how these extragalactic  GMCs are formed remains an intriguing open question to be answered by future studies.

The authors would like to thank Jean Brodie for pointing us to NGC 5253 and its accretion streamer and to Cody Raskin and Mark Richardson for useful conversations. We acknowledge  support from NASA under theory Grant No. NNX09AD106  and from the National Science Foundation under grant AST 11-03608.  All simulations were conducted on the ``Saguaro" cluster at the Arizona State University Advanced Computing Center, using the FLASH code, a product of the DOE ASC/Alliances funded Center for Astrophysical Thermonuclear Flashes at the University of Chicago. Figures~\ref{init}-\ref{dggsum} were created using the {$\bf{yt}$} analysis package (Turk \etal 2011). This work performed under the auspices of the U.S. Department of Energy by Lawrence Livermore National Laboratory under Contract DE-AC52-07NA27344.

{}
\end{document}